\documentclass[%
 aip,
 amsmath,amssymb,
 reprint,
 nofootinbib,
 floatfix,
]{revtex4-1}

\usepackage{graphicx}
\usepackage{dcolumn}
\usepackage{bm}
\usepackage[utf8]{inputenc}
\usepackage[T1]{fontenc}
\usepackage{mathptmx}
\usepackage{etoolbox}
\usepackage[dvipsnames]{xcolor}
\usepackage{soul}
\usepackage[T1]{fontenc}
\usepackage{float}
\usepackage{nomencl}
\usepackage{hyperref}

\makeatletter
\def\@email#1#2{%
 \endgroup
 \patchcmd{\titleblock@produce}
  {\frontmatter@RRAPformat}
  {\frontmatter@RRAPformat{\produce@RRAP{*#1\href{mailto:#2}{#2}}}\frontmatter@RRAPformat}
  {}{}
}%
\makeatother

\begin{document}

\preprint{AIP/123-QED}

\title[]{Control of bow shock induced three-dimensional separation using bleed through holes}
\author{Hemanth Chandravamsi}
\author{Sourabh Bhardwaj}
\author{K. Ramachandra}
\author{R. Sriram}
\affiliation{Department of Aerospace Engineering, Indian Institute of Technology Madras, India}

\date{\today}

\begin{abstract}
The unsteady three-dimensional separated flow on a wall induced by a square protrusion (approximately twice the local boundary layer thickness in width and height), and its control by means of passive suction through holes, is investigated using wind tunnel experiments at Mach $2.87$. The baseline flow without any control was characterized and compared against the cases with bleed. A bow-shaped separation line on the wall with a mid-span separation length of $5.57\delta$ from protrusion face was traced from oil-flow visualization. The averaged pressure distribution surveyed using static pressure ports placed on the wall has mapped plateau, high-pressure, and a low-pressure region in the separated flow, distinctive to three-dimensional interactions. Ten control configurations were tested with suction holes placed along mid-span in the different pressure zones. Significant spanwise `Mean Reduction in Separation Length' of up to $0.93\delta$ was observed from oil-flow visualization. A comparison of observations from various control configurations suggested that bleeding the flow from the high-pressure region could in general delay the separation and reduce the bubble size. Further, time-resolved schlieren visualizations have confirmed reduction in both `mid-span separation length' and `shock-intermittent-region' with the introduction of suction in high-pressure region. Fourier and Proper Orthogonal Decomposition analysis done on the schlieren data has confirmed the presence of low-frequency separation-shock oscillations at Strouhal Numbers of order $10^{-2}$, both with and without control. Furthermore, the amplitudes of separation-shock oscillations in the spectrum were reduced with the introduction of suction simultaneously from two holes placed in high and low-pressure regions.
\end{abstract}

\maketitle

\section{Introduction}

Unsteady separated flows induced due to Shock Wave Boundary Layer Interactions (SBLI) can cause violent thermo-mechanical loads which are capable of causing catastrophic damage to the components of supersonic/hypersonic systems such as intakes, isolators and control surfaces. With seventy years of history on SBLI research and the recent advancements in diagnostics and computational facilities, various aspects of turbulent SBLI due to nominally two-dimensional geometric configurations including the ``low-frequency unsteadiness” were well resolved \cite{clemens2014low}.

Several studies from the past few decades have established how various flow parameters such as shock strength, Mach number ($M$), Reynolds number ($Re$), and boundary layer profile affect SBLI in general, as well as the induced flow separation \cite{DeleryChapter}. In addition, the unsteady aspects and the arguments supporting upstream and downstream influences are also studied extensively using both experimental \cite{ganapathisubramani2007effects,dussauge2006unsteadiness}, and computational approaches \cite{pirozzoli2006direct}. All these studies together establish a good understanding of the physics of two-dimensional SBLI. However, in several practical engineering applications, three-dimensional interactions are often encountered.

Although there are reports in the literature addressing three-dimensional SBLI, each of them was configuration specific, lacking any universal picture. \citet{westkaemper1968turbulent} and \citet{voitenko1966supersonic} in 1966 have independently reported the features of three-dimensional separated flowfields induced by cylindrical protrusions placed in supersonic flow. Subsequently in the following years, extensive experimental \cite{zheltovodov2011ideal} and a few RANS based computational studies \cite{schmisseur2000exploration,zheltovodov2000verification} were performed on three-dimensional configurations such as sharp unswept fins \cite{zheltovodov1982regimes}, sharp swept fins \cite{zheltovodov1986three}, semi-cones\cite{Semicone} and swept compression ramps \cite{settles1980investigation}. The nature of shock topology, streamline patterns in the vicinity of the object, pressure distribution and important flow transitions under varying geometric and flow parameters were studied in the above mentioned investigations. In the recent past, \citet{mowatt2011three} have investigated the interaction of three-dimensional curved shock and boundary layer on a variety of curved surfaces using a set of experimental and RANS based simulations. Pickles et al. \cite{pickles2019mean} in their recent study have investigated three-dimensional SBLI under fin-on-cylinder configuration. Their experimental and computational studies have revealed the topological features of the separated flowfield along with the extent of three-dimensional relief offered by the cylinder’s lateral surface under various fin-angles, inflow Mach numbers and cylinder diameters.

With only a few reported investigations on such diverse configurations, possible universal aspects concerning three-dimensional interactions are not yet understood, even for classes of problems such as the SBLI due to bow shock. Further, owing to the challenges in terms of computational cost to perform high fidelity numerical studies using Large Eddy Simulations (LES) and Direct Numerical Simulations (DNS), and limited three-dimensional characterization capability of experimental flow diagnostics, researchers in the past have largely managed to capture only the time averaged picture of the interactions. Recent studies by \citet{SB2022} was the first to propose a general scaling law (valid for various protuberance shapes and sizes as well as supersonic freestream Mach numbers) relating separation length in protuberance induced SBLI with the inviscid bow shock's radius of curvature and freestream conditions. However, a comprehensive understanding of the unsteady aspects of these interactions is not well established in the literature. With relatively sparse knowledge on unsteadiness, it is also difficult to evaluate the critical aerodynamic and thermal loads associated with three-dimensional interactions, which are essential in designing efficient and safer high-speed flights.

Characterizing the interaction dynamics and underlying mechanisms is also essential to implement control mechanisms to overcome the adverse effects. Various active and passive flow control techniques for unsteady SBLI have been explored and demonstrated successfully for nominally two-dimensional configurations. Passive control devices, primarily due to their simplicity in implementation and other essential benefits, have attracted several researchers and engineers as potential candidates for mitigating separation and unsteadiness. Among the various passive control methods, boundary layer suction/bleed has received significant attention \cite{raghunathan1988passive,delery1985shock} in all speed regimes. In this technique, fluid close to the wall, either upstream of the separation or inside the bubble, is removed using suction, taking out the local low momentum fluid.

The bleed configurations can be classified into three categories based on the vent design: slots/perforations/holes. The use of these devices is not only limited to the control of SBLI but has also been widely applied to control other flow fields with separation and unsteadiness. For instance, in the context of supersonic intake flow control, \citet{herrmann2011experimental} and \citet{soltani2015effects} have used bleed through slots to control the `intake buzz.' With regard to two-dimensional-SBLI too, slot control has attracted significant attention, especially since slots can be in nominally two-dimensional configurations. \citet{krogmann1985effects}, and \citet{hamed1995shock} have used slot(s) oriented normal to the primary flow direction producing a two-dimensional control effect for interactions in transonic and supersonic regimes respectively. With the former being experimental and the latter numerical, both studies report the effect of various bleed parameters in controlling the two-dimensional SBLI. Contrary to placing slots normal to the flow, studies were also performed with slots oriented along streamwise direction, essentially inducing a three-dimensional control effect. For instance, \citet{smith2003experimental} and \citet{holden2005separated} have used a series of streamwise bleed slots to delay the onset and reduce the extent of shock induced separation. Interestingly, the study by \citet{holden2005separated} notes that although the streamwise slot configuration they employed is three-dimensional, the control effect induced on the interaction stands relatively two-dimensional (meaning the controlled flowfield remained nearly uniform along the spanwise direction).

Perforated strips were also employed by researchers to control SBLIs. \citet{thiede1984active} have experimentally shown that perforated strips can delay shock induced separation and stabilize the shock motion, improving the airfoil characteristics on which these interactions occur. \citet{jegadheeswaran2019perforated} have investigated the effect of a perforated wall in scramjet engine setup through two-dimensional numerical simulations. They report that the bleed from perforations causes the size of the separation bubble to shrink by inducing a feedback loop between the upstream and downstream regions of the separation bubble. However, such nominally two-dimensional control configurations may not be effective for three-dimensional flowfields.

%\citet{sriram2014shock} have performed shock tunnel experiments at Mach 5.96, demonstrating the control of impinging shock induced separation using boundary layer bleed through closely spaced spanwise rows of holes. Their setup mimics a quasi-two-dimensional configuration with the array of small holes acting as perforations. However, extending such two-dimensional demonstrations to three-dimensional configurations and the potential complications that may arise still need to be addressed.

Alongside slots and perforations, holes were also employed to control SBLI. However, holes can be placed in different arrangements. Their number, size, and relative location to the interaction region can play an important role. One simple arrangement widely used to control two-dimensional interactions is to have a uniformly spaced array of holes that produces a global two-dimensional effect. A few applications that employ this kind of arrangement include \citet{ghosh2010simulation} and \citet{willis1995flowfield}, in which they control separation due to an impinging shock boundary layer interaction. In the hypersonic regime, \citet{sriram2014shock} have performed shock tunnel experiments to control impinging shock induced separation through an array of spanwise bleed holes. Considering the small size of the bleed holes and the spacing between them in their experiments, their bleed set up can be categorized to fall between perforations and hole based control. However, unlike most slot and perforation configurations, a bleed hole that is sufficiently distanced from other bleed holes induces a three-dimensional effect. Literature on hole control suggests that such three-dimensional bleed hole configurations can be used to control two-dimensional interactions. \citet{schoenenberger1999flow}, and \citet{bodner1996experimental} have investigated the effect of a single bleed hole on the development of turbulent boundary layer downstream without the presence of any shock or SBLI. \citet{rimlinger1992three} (1992) have numerically studied the effect of a single bleed hole and its position relative to impinging shock and demonstrated its control effect. The single bleed hole was noted to result in three-dimensional control effect in their flow field. The same group, later in 1996, extended their study \cite{rimlinger1996shock} and explored the effect of rows of multiple holes arranged three-dimensionally in a staggered fashion. They report that when the holes are placed sufficiently upstream of the incident shock, the bleed effect can adequately block the shock-induced adverse pressure gradient from propagating further upstream, thereby controlling separation. Such studies tell us that even in nominally two-dimensional SBLIs, the three-dimensional effect from bleed holes can induce a good control effect. While carrying out the present study, we have hypothesized that the same three-dimensional control effect from the holes can also be used for controlling three-dimensional interactions, possibly with better efficiency. Furthermore, in the context of three-dimensional interactions, the placement of hole(s) can even be critical. To the best of our knowledge, there is no detailed study in the literature concerning the bleed control of three-dimensional shock induced separation due to protrusion.

% While carrying out the present study, we have thought that, 
% The study of \citet{ghosh2010simulation} reports that, under certain bleed rates, the separation region gets eliminated, and the bleed induces local pressure disturbances near the wall. 

The research presented in the current article is a part of the broad study initiated at the Department of Aerospace Engineering, IIT Madras to understand and address the issues concerning three-dimensional SBLI. The current study concerns the characterization of three-dimensional shock boundary layer interaction induced by a square protrusion placed in a supersonic flow (of nominal Mach number 2.93) and a demonstration of passive suction based strategic control to mitigate separation and flow unsteadiness. This canonical three-dimensional configuration of the square protrusion (with its face oriented perpendicular to incoming flow) was especially chosen for the current study to explore the implications of the interaction between strong detached shock (generated due to the bluntness of protrusion) and turbulent boundary layer. The results from the study are thus qualitatively applicable to any bow shock induced separation, such as that in blunt fin induced SBLI. 

The paper is organized as follows. After the present section, the experimental setup designed to carry out the current study, i.e., the wind tunnel, flow diagnostics, and control setup, shall be detailed in section \ref{sec:setup}. Following that, in section \ref{sec:baseline}, the mean and unsteady nature of the baseline\footnote{The term `baseline' in this paper is used to refer to the flow configuration without any control.} case i.e., the protrusion induced three-dimensional separated flowfield, is discussed. Then, in sections \ref{sec:control1} and \ref{sec:control2}, the effect of bleed holes placed inside the separated region under various configurations (designed based on baseline results) is discussed. In particular, tools such as Fourier analysis and snapshot Proper Orthogonal Decomposition (POD) were employed to get a deeper understanding of the unsteady dynamics of the baseline flowfield and the control effect. Finally, the conclusions are laid out.

% The following two sections \ref{sec:baseline} and \ref{sec:control1} are devoted to present the results corresponding to baseline\footnote{The term `baseline' in this paper is used to refer to the flow configuration without any control.} flow and the effect of bleed holes on the flowfield respectively.

\section{Experimental set-up and flow diagnostics} \label{sec:setup}
\subsection{Wind tunnel facility and model set-up}

\begin{figure}[htpb]
    \centering
    \includegraphics[width=85mm]{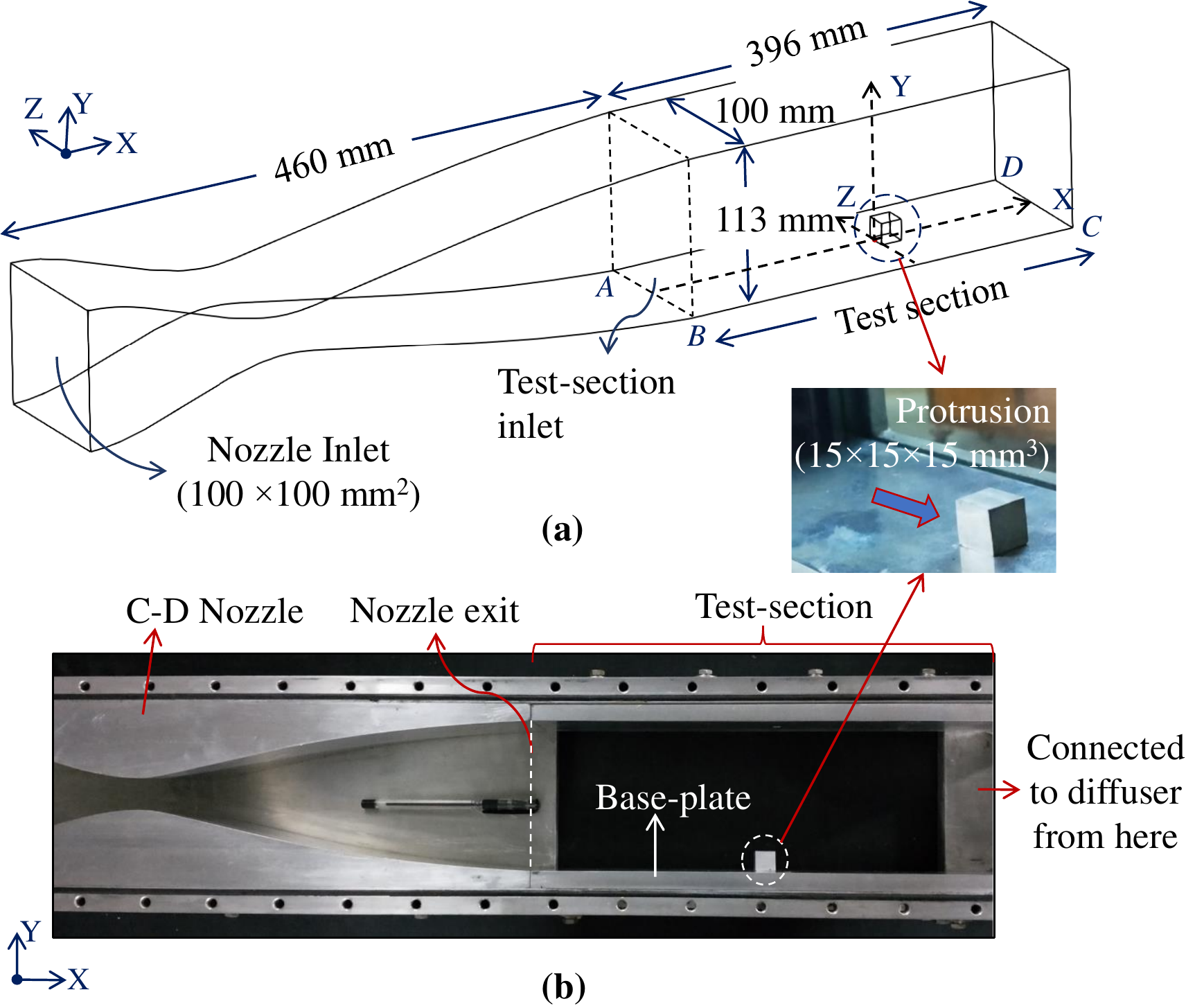}
    \caption{(a) Schematic of the wind tunnel designed to study protrusion induced three-dimensional shock boundary layer interaction. (b)	Cross-sectional view of nozzle and test-section (pen placed as a reference for scale)}
    \label{setup1}
\end{figure}

Experiments were performed in the supersonic blowdown wind tunnel at the Gas Dynamics Laboratory, IIT Madras (for details, refer to \cite{SB2022}). The tunnel is being operated at a nominal stagnation pressure ($P_o$) and stagnation temperature ($T_o$) of $6 \pm 1.1\%$ bar (absolute) and $300\pm 1\%$ K, respectively, with each run lasting nearly $10$ seconds. The schematic of the nozzle and test-section assembly designed for the current study is shown in Fig \ref{setup1}. The set up includes a 396 mm long test-section with a rectangular cross-section of span and height, $100$ mm and $113$ mm, respectively, with a provision of optical access on the side walls (for visual inspection and schlieren visualization). This test-section is attached to the exit of a contoured supersonic nozzle of exit to throat area ratio of $3.96$. The schematic of the set up with origin (origin is at the spanwise center, on wall protrusion junction) and the coordinate axis orientation can be understood from Fig \ref{setup1}a. In order to make alterations as well as to enable taking it out to capture the photograph of surface streakline pattern after the experimental run, the base-plate (described by plane ABCD in Fig \ref{setup1}a) was designed in a modular fashion to be easily removable from the test section while maintaining the desired geometrical integrity of the whole set up.

\begin{figure}[htpb]
    \centering
    \includegraphics[width=65mm]{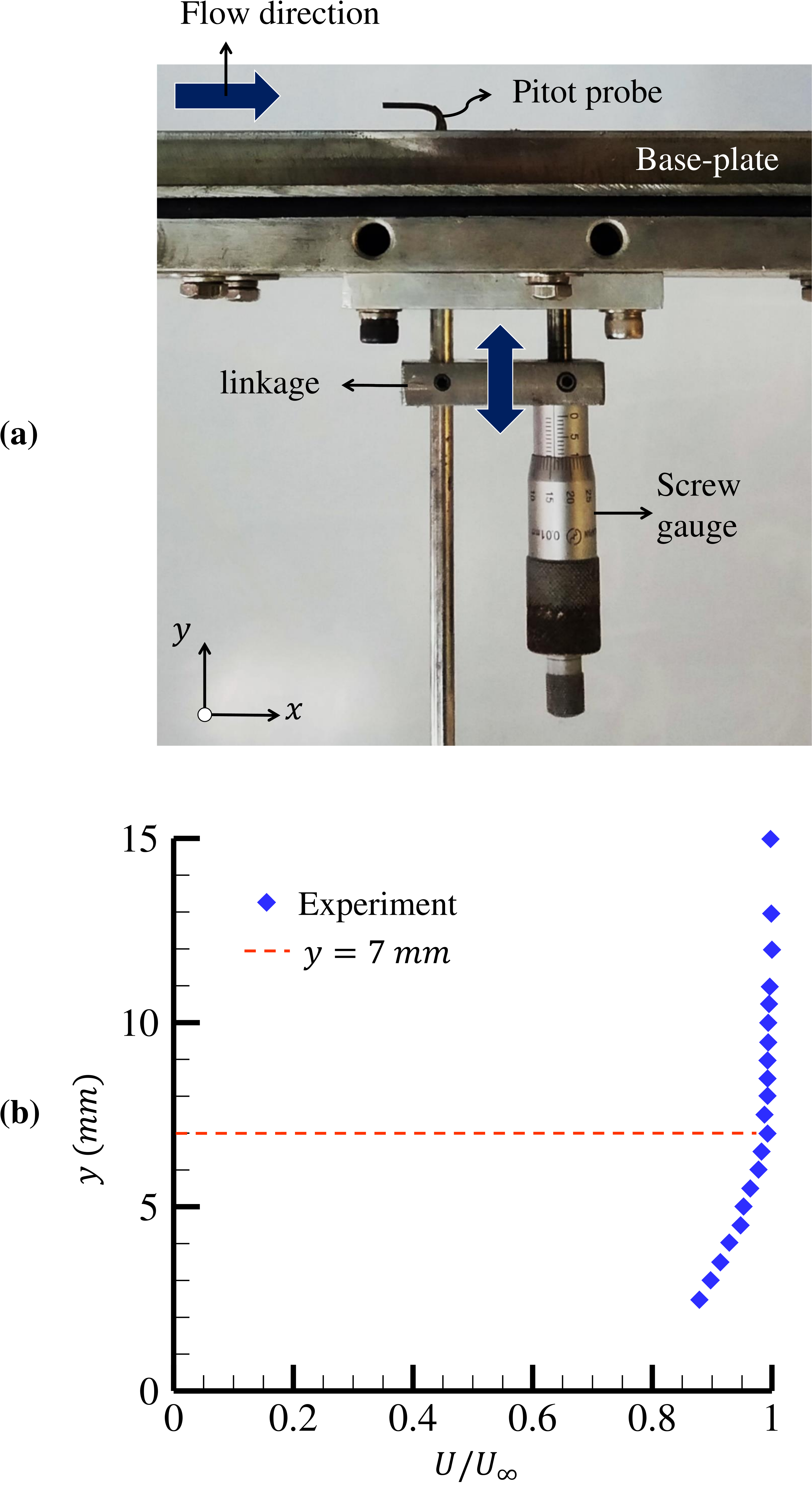}
    \caption{(a) Precisely moving pitot probe assembly used to survey the boundary layer and freestream of the wind tunnel mounted at protrusion placement spot. (b) Boundary layer profile probed at protrusion mounting location using the precisely moving pitot-probe arrangement (without any obstacle/model being placed in it).}
    \label{setup2}
\end{figure}

The freestream conditions and boundary layer were characterized using a precisely moving pitot-probe arrangement shown in Fig \ref{setup2}a. The internal and external diameters of the pitot probe are $0.9$ mm and $1.8$ mm respectively. The freestream Mach number in the test section was $2.87 \pm 4.4\%$. A summary of the pitot survey results is presented in Fig. \ref{setup2}a and table \ref{Pitot-table}. The pitot survey in the test section (without mounting the protuberance) was done at a distance of $170$ mm from the nozzle exit, which is also where the protrusion's face was mounted. The time-averaged velocity profile obtained from the pitot survey is shown in  Fig \ref{setup2}b, from which the boundary layer thickness ($\delta$)  was measured to be $7$ mm. Based on boundary layer thickness, the protrusion’s height ($h$) and width were chosen to be roughly two times $\delta$ ($15$ mm), with an overall dimension of $15 \times 15 \times 15$ mm$^3$ (The protrusion can be seen in Fig \ref{setup1}). The dimensions of the cube were so chosen to facilitate the comparison of the current study's results with the corresponding two-dimensional case (forward facing step) study conducted by \citet{murugan2016shock}. The step height in their case was taken as two times their incoming boundary layer thickness. The protrusion was mounted at the spanwise center, with one of its face perpendicular to the streamwise direction.

\begin{table*}[htpb]
\caption{\label{Pitot-table}Free stream conditions inside the test-section and boundary layer thickness at protrusion placement spot}
\begin{ruledtabular}
\begin{tabular}{@{}cccccl@{}}
\textbf{$M_\infty$} & \textbf{$Re_\infty$}      & \textbf{$U_\infty$} & \textbf{$P_\infty$} & \textbf{$T_\infty$} & \textbf{$\delta$} \\
$2.87\pm 4.4\%$       & $40.4 \times 10^{6}$ 1/m & $612 \pm 1.2\%$ m/s  & $19200 \pm 0.05\%$ Pa   & $114 \pm 1\%$ K      & $7$ mm
\end{tabular}
\end{ruledtabular}
\end{table*}

\subsection{Flow diagnostics}

Oil-flow visualization, schlieren imaging, and surface pressure measurements (averaged) were the flow diagnostics used in the current study. The details of the diagnostics are presented in the paper by \citet{SB2022}. Oil-flow surface streakline visualization was performed on the base plate where the protrusion was mounted for both baseline and control cases to highlight important features such as the separation line and streakline pattern inside the separation bubble. A mixture of Titanium Dioxide (TiO2) and SAE-$30$ grade mineral oil with Oleic acid as the emulsifying agent was prepared and spattered in the region of study to carry out oil-flow experiments \cite{Oilflow}. The steady-state streakline pattern generated at the end of each experiment is captured using a camera. The geometry of the separation line is also recorded for comparison and analysis purposes by means of grid lines drawn on the surface. Multiple experiments were performed to check the repeatability of the results and found out to be so.

Schlieren visualization with light passing perpendicular to the side walls was used to observe the separated flow field's shock patterns and features. The high-speed ‘Photron FASTCAM SA4, Model 500K-M1’ camera was used to capture the images. With a time resolution of $30,000$ frames per second, schlieren visualizations were also performed to characterize the shock motion and unsteadiness quantitatively.

The time-averaged surface pressure distribution on the bottom wall in the separated flow field was surveyed by means of pressure ports and an electronic pressure scanner (Scanivalve DSA-3217). For this purpose, a layout of thirty holes (static pressure taps) of $0.6$ mm diameter each was made on the base plate with the help of insights gained from baseline oil-flow experiments. The layout of the ports is designed in a fashion to understand the pressure distribution and identify various pressure zones corresponding to the interaction. In addition, readings obtained from a set of four experiments were considered to acquire a converged set of average pressure values.

\subsection{Control set up}

\begin{figure}[htpb]
    \centering
    \includegraphics[width=70mm]{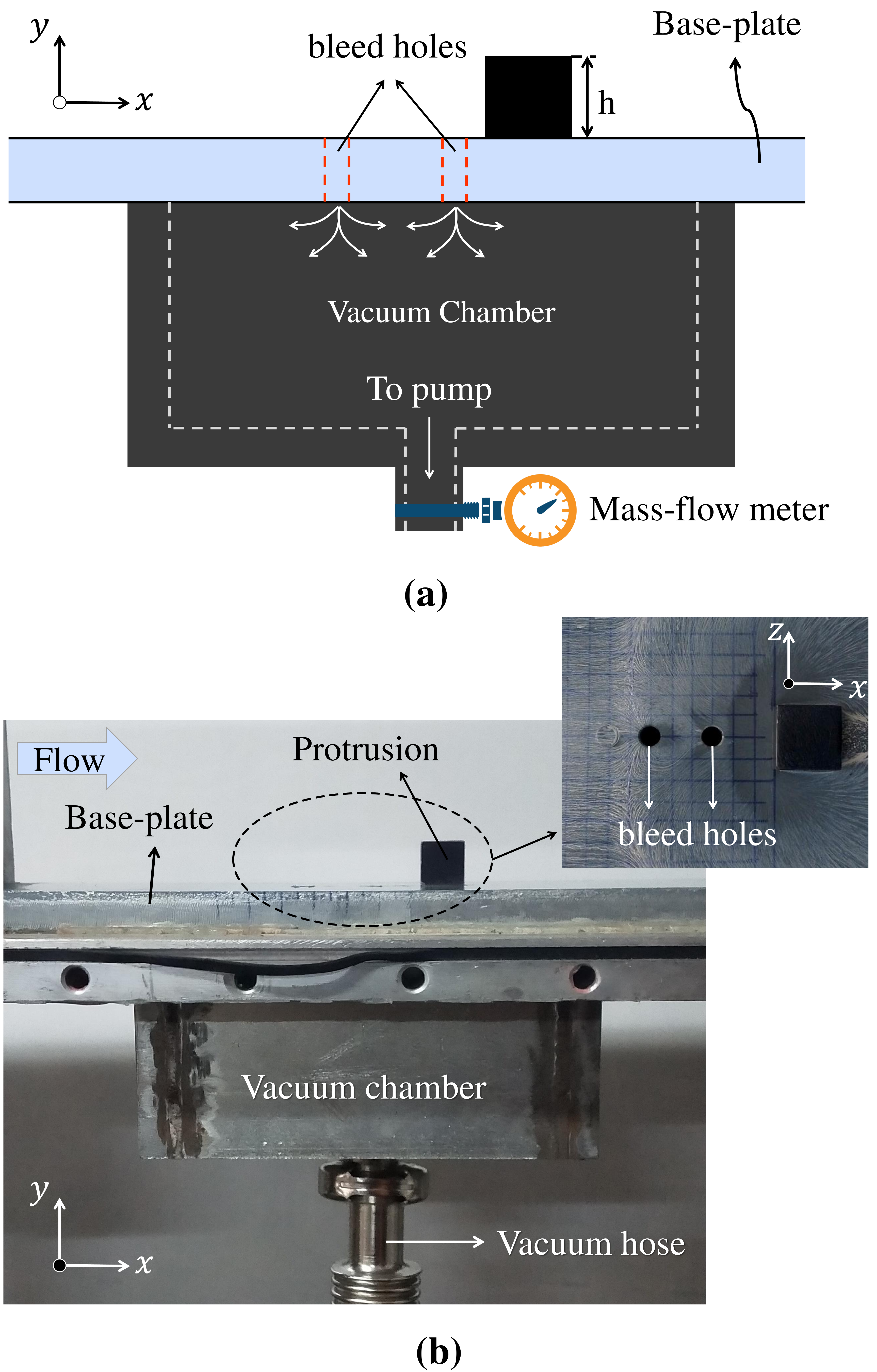}
    \caption{Suction chamber and baseplate assembly for control experiments: (a) Schematic sketch (b) Photograph of the set-up.}
    \label{setup3}
\end{figure}

Passive suction is implemented in the separated flow region through circular holes along the centerline on the bottom surface. The control experiments were designed based on the insights gained from the surface static pressure measurements obtained on the bottom surface for the baseline case. Based on separation features and pressure distribution characterized by baseline experiments, a total of ten bleed-hole configurations were designed to understand the effect of suction from various pressure zones present in the separated region. To set up various control configurations, two base-plates were prepared with three bleed holes of $5$ mm diameter on each of them. Using two plates enabled the placing of holes at locations that were only slightly off from locations in the first plate, so as to see if small changes in hole placements may vary the flow drastically.  Different control configurations were achieved by strategically closing some holes and keeping the others open. To close the holes, a hard cylindrical rubber of a slightly larger diameter than the hole was passed through and flush mounted into the holes with a flat cut on the plate surface. The specific details of the geometric positioning of the bleed holes for various configurations shall be discussed in section \ref{sec:control1} after presenting the data of surface static pressure.

To draw the fluid out from these bleed holes (each of 5 mm diameter), a rectangular suction-chamber of dimensions $149 \times 64 \times 60$ mm$^3$ was placed beneath the holes under the bottom surface of the test-section as shown in Fig \ref{setup3}. The bleed holes drilled on the test-section’s baseplate were exposed to the vacuum chamber’s suction from beneath. The chamber is evacuated by an external vacuum pump which generates sufficient vacuum to choke all the bleed holes.

It is reasonable to expect that the bleed mass flow rate and the extent of local flow field manipulation it results in are correlated. Thus it is of interest to measure the bleed mass flow rate of the different control configurations. It is worth noting that the bleed rate can vary from one configuration to another depending on the location of the bleed hole(s) in different pressure zones. Therefore, in an attempt to understand the correlation between the bleed mass-flow, hole positions and control achieved, mass-flow rate measurements were performed for some critical control cases using a `Bronkhorst D-6380-DR' mass-flow meter. The mass-flow meter was placed after the vacuum chamber before the bleed air enters into the vacuum pump. The mass-flow rate readings presented are accurate to within $\pm 0.5 \%$ kg/hr. The experimental results shall be discussed in the following section.

%%%%%% SECTION-2

\section{Three dimensional shock induced separated flowfield} \label{sec:baseline}

The time-averaged features of the flowfield due to a square protrusion under the same operating conditions as in the present work were detailed in the recent paper by \citet{SB2022}. In this section, we revisit some of the key aspects of time-averaged flowfield to establish the nomenclature, concepts required for further discussion, and the rationale behind the control strategy drawn from various pressure regions. Additionally, this section also presents a characterization of unsteadiness in the flowfield based on time-resolved schlieren. These aspects of the baseline case are subsequently used for comparison with the control cases.

\subsection{Oil flow visualization}

\begin{figure}[htpb]
    \centering
    \includegraphics[width=65mm]{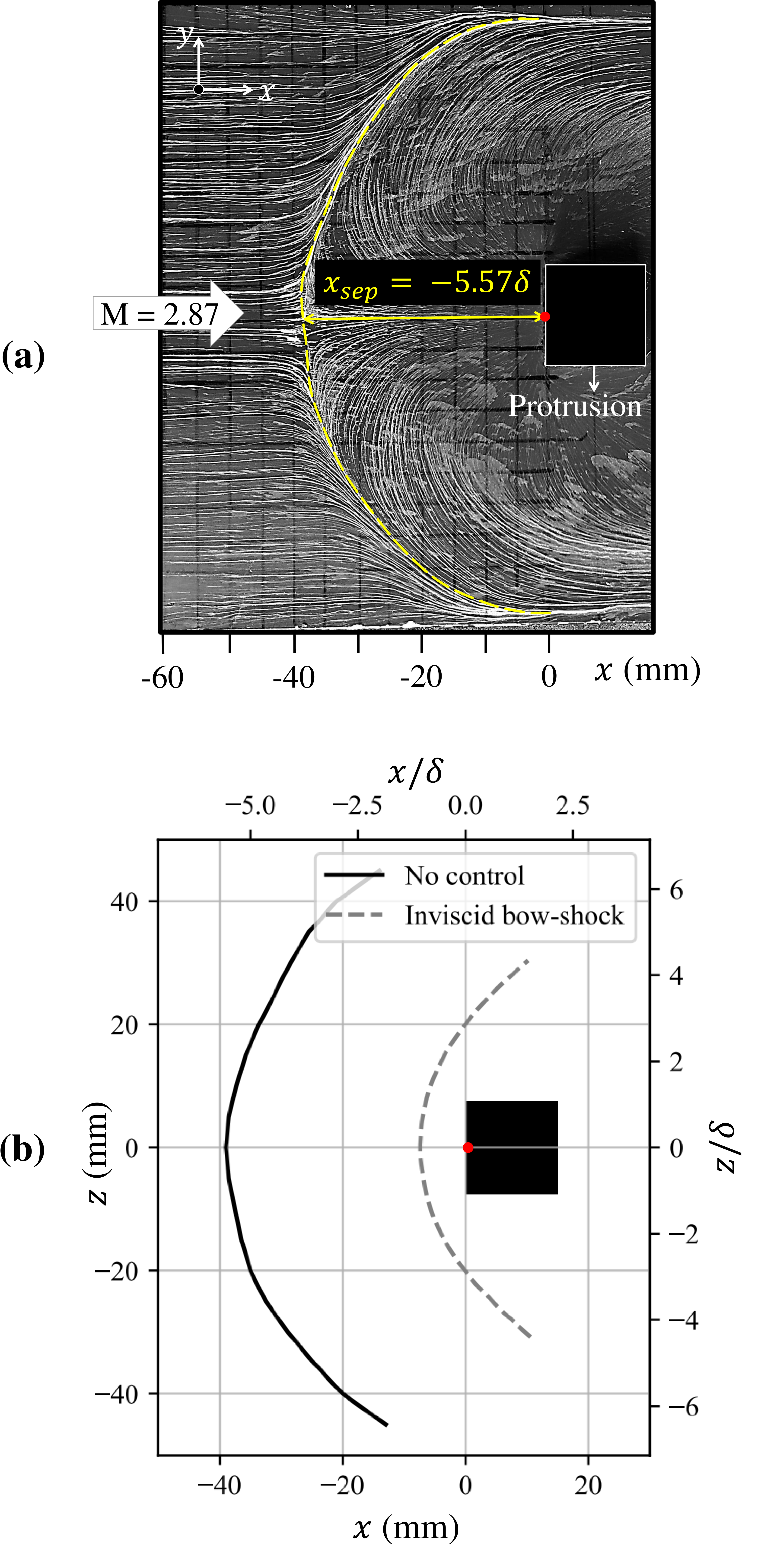}
    \caption{(a) Oil-flow surface streakline pattern for the baseline case with separation line highlighted, (b) Separation line position for the baseline case (solid black line) traced from oil-flow visualization, and Inviscid bow-shock position from numerical Euler simulations}
    \label{oilflow}
\end{figure}

Oil flow visualization for the baseline case reveals the streakline pattern on the base-plate, highlighting the flow separation line and the streakline pattern inside the separation bubble, as shown in the Fig \ref{oilflow}a. The separation line can be identified as a bow shaped thin region of convergence of the streak lines between the incoming flow and the separation bubble. The separation length, which is defined as the distance between the separation line and the protrusion surface at the spanwise centre, was measured to be $5.57\delta$ for the square-faced protrusion as indicated in the figure. With the help of grid markings made on the bottom surface, the separation line is traced and plotted as shown in Fig \ref{oilflow}b on the right (in solid-black line). The bow-shock position on bottom wall corresponding to inviscid flow over this protrusion was also represented in the Fig \ref{oilflow}b for comparison (this bow-shock’s position was found through numerical Euler simulation, as discussed in \citet{SB2022}). It is worth noting that, due to the interaction, the (separation) shock has moved upstream by an amount of $5.3$ times the stand-off distance of the inviscid bow-shock along midspan. Also, the current midspan separation length can be noted to be roughly $1.6$ times lower than the separation length ($8.2\delta$) reported for the corresponding two-dimensional case (forward facing step) by \citet{murugan2016shock}.

\subsection{Surface static pressure measurements}

\begin{figure*}[hbpt]
    \centering
    \includegraphics[width=145mm]{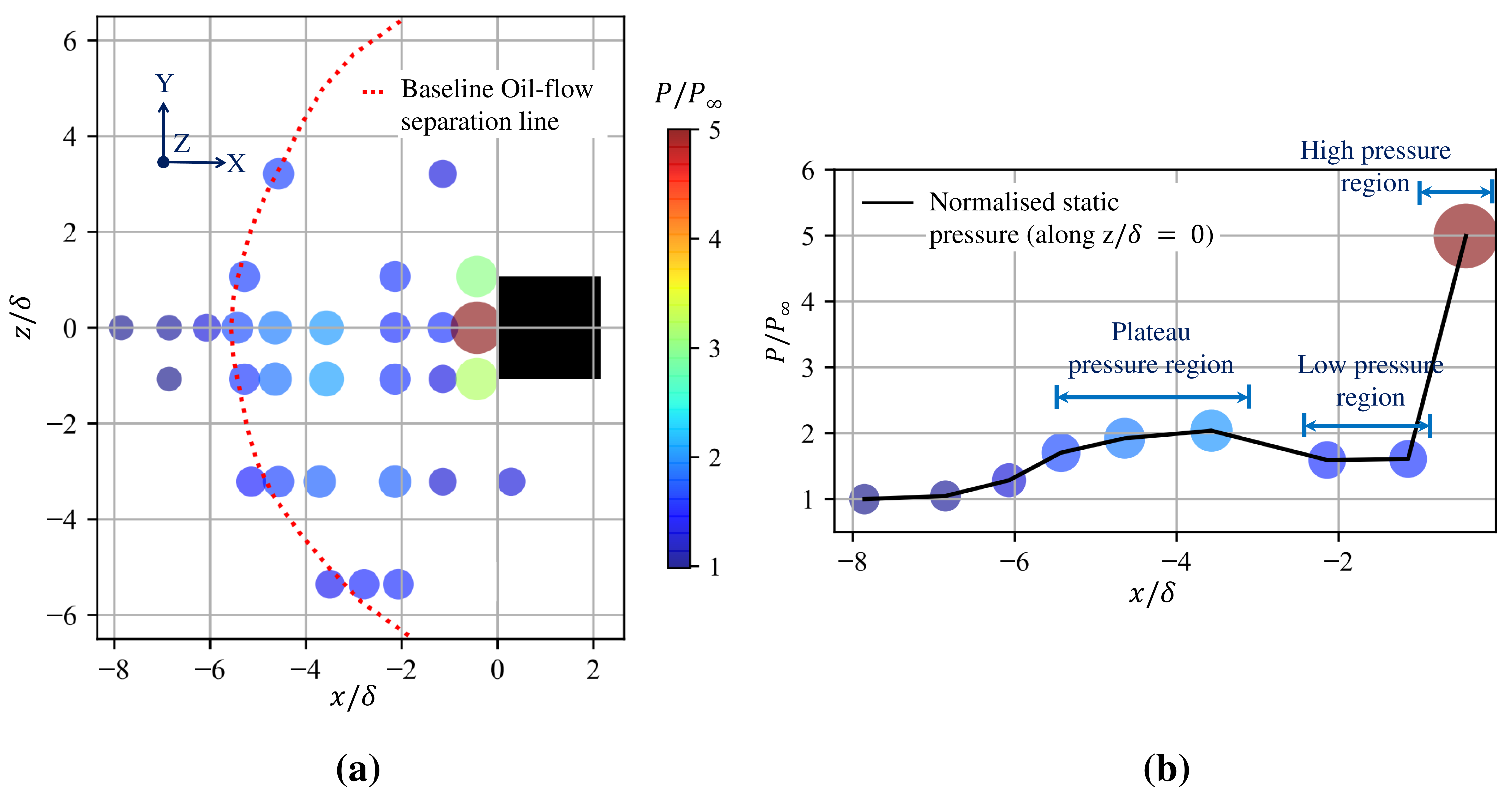}
    \caption{(a) Surface pressures at various locations on bottom surface measured through the surface static pressure taps. The centres of the circle are the positions of the static pressure taps. Colour and radius of each solid circle represents the magnitude of pressure. (b) Surface pressure distribution along midspan ($z/\delta = 0$). Various pressure regions are indicated.}
    \label{Static_P}
\end{figure*}

With the reference of the separation line traced from oil-flow visualization, the layout for static pressure ports was designed as shown in Fig \ref{Static_P}a (each circle represents a sensor there, with its radius and colour indicative of the magnitude of local pressure). It can be seen that the line along spanwise centre was particularly resolved to have a total of 9 ports. In addition to that, at a few streamwise locations, taps were placed symmetrically on either side of the centreline to ensure symmetry of the flow. Otherwise, most of the taps were concentrated on the left side of the centreline as seen in Fig \ref{Static_P}a (bottom portion in the figure), expecting symmetry in the field (as it turned out to be from measurements). It can also be seen that along the separation line there are a total of six pressure taps. Along the spanwise centreline, it can be observed that the first two taps measure freestream pressure. In the third sensor, which is a little upstream of the mean separation line, the pressure starts to rise. This is attributed to the upstream influence, due to which the pressure starts to rise before the separation. Following this, there is a region of plateau pressure with a value close to twice the freestream pressure inside the separation bubble up to a distance of around $4\delta$ from the protrusion. Downstream of this plateau region, the surface pressure drops, reaching values as low as $1.5$ times freestream pressure between the distances $\delta$ and $2\delta$ away from the protrusion. This local region of low pressure in the protrusion induced separation bubble is due to the three-dimensional relieving effect, and such a region is not observed in two-dimensional SBLI in which the plateau pressure inside the bubble is directly followed by the pressure rise at reattachment. The relieving in protrusion induced SBLI is due to the spanwise turning of streamlines inside the separation bubble which is evident from the surface oil-flow streakline patterns presented in Fig \ref{oilflow}a. It is in this region where the relieving begins that the pressures are low. However, very close to the protrusion, the pressures are very high (5 times the freestream pressure) due to reattachment and stagnation of flow at the protrusion front face. The trend is observed even along lines that are close to the spanwise centre line, though the values of the peak pressure are observed to be lower. A similar pressure profile with pressure dip was also reported by \citet{schulein2001documentation} in 2001 in their study on three-dimensional separation due to fins (see figures 5.17a and 5.22 in the book by \citet{babinsky2011shock}). In the neighbourhood of the spanwise centreline, based on the surface pressure distribution, three distinct regions can be demarcated, viz., plateau region, low-pressure zone and high-pressure zone. Far away from the centreline, the plateau pressure after the separation is followed by a pressure drop, and no high pressure zones are observed. The high-pressure is thus confined to a width comparable to the width of the protrusion. Based on the pressure distribution along the spanwise centreline, control experiments were designed to explore the effect of suction from various pressure regions on the separation length and unsteadiness.

\subsection{Schlieren flow visualization} \label{sec:schlieren1}

\begin{figure}[htpb]
    \centering
    \includegraphics[width=70mm]{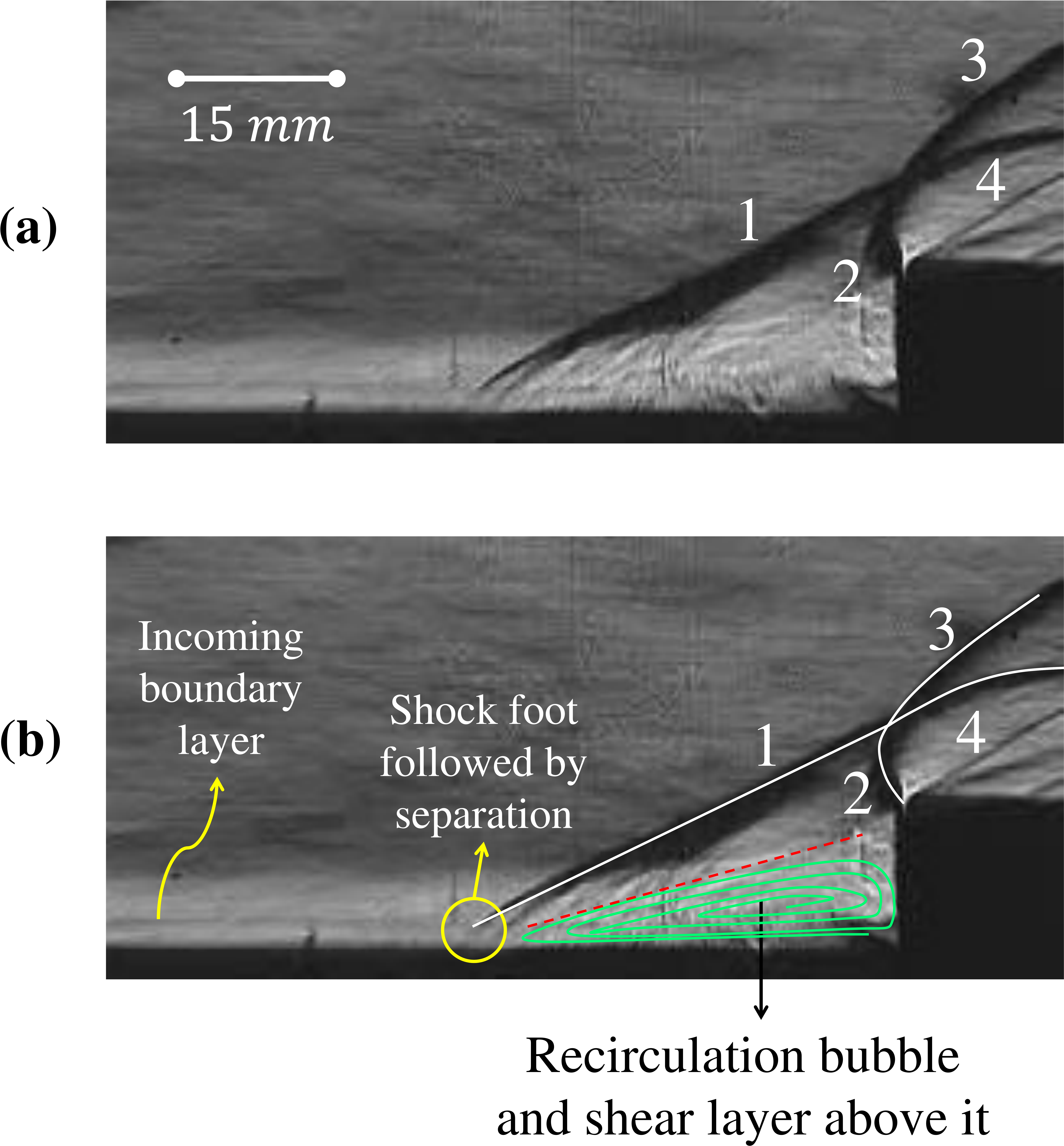}
    \caption{(a) An instantaneous schlieren snapshot taken for the baseline case (raw image), (b)	Various features of the three-dimensional shock boundary layer interaction flow field labelled.}
    \label{schlieren1}
\end{figure}

The Schlieren visualization (Fig. \ref{schlieren1}a) reveals various features present in the flowfield such as: incoming boundary layer, separation shock, reattachment shock, re-circulation bubble and shear layer as labelled in Fig \ref{schlieren1}b. As shown in the Fig. \ref{schlieren1}, in the spanwise central plane, type-6 Edney shock interaction \cite{edney1968anomalous} was observed between separation shock (labelled with number 1) and reattachment shock (labelled with number 2). The interaction of these two shocks resulted in two reflected shocks labelled with numbers 3 and 4 Fig. \ref{schlieren1}.

\begin{figure}[htpb]
    \centering
    \includegraphics[width=85mm]{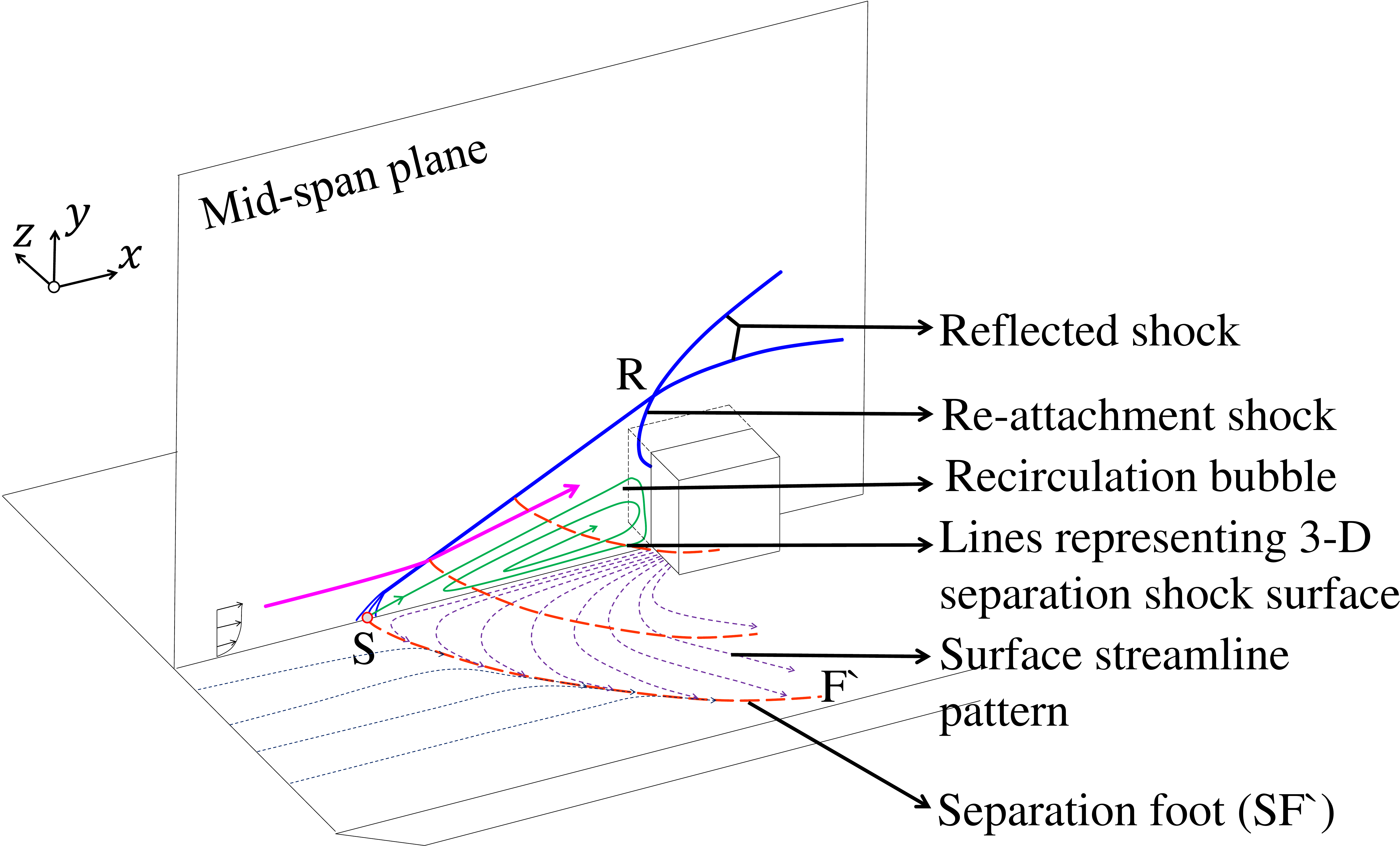}
    \caption{Schematic of three-dimensional separation in front of square protrusion.}
    \label{schematic}
\end{figure}

Combining the oil-flow data and the schlieren visualizations, a schematic of the three-dimensional separated flow upstream of the square protrusion is shown in Fig \ref{schematic}. The schematic shows important features such as the three-dimensional separation shock topology (represented with red dashed lines), the separation line (SF'), the surface streakline pattern, shock interaction in the midspan plane, and the `recirculation' bubble inside the separation zone in the midspan plane. The bubble is not shown to have closed streamlines in the schematic (in contrast to the two-dimensional bubbles) since the streamlines spiral towards a core and subsequently turn spanwise, resulting in a horseshoe vortex (detailed topology is discussed in \citet{SB2022}, 2022).

\begin{figure}[htpb]
    \centering
    \includegraphics[width=85mm]{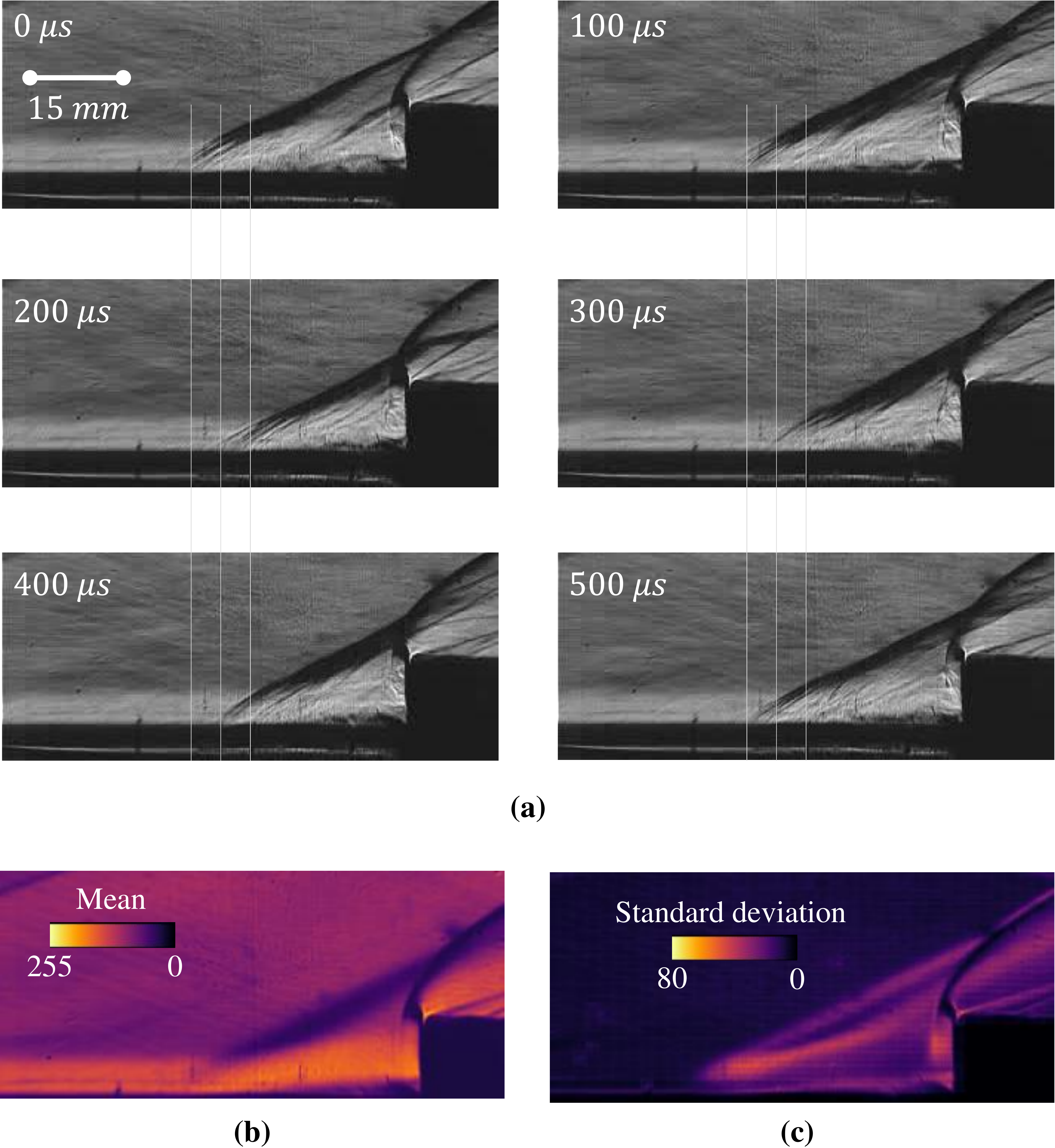}
    \caption{(a) Instantaneous schlieren visualization snapshots (without control). (b) Mean and (c) Standard deviation of pixel intensity of schlieren images}
    \label{schlieren2}
\end{figure}

With the time-resolved schlieren imaging, the positions of separation shock and reattachment shock were seen to vary temporally. Some instantaneous schlieren snapshots taken during the run are shown in the Fig. \ref{schlieren2}a. The shock position and separation bubble size can be clearly seen to vary over time in these images. For instance, this difference can be clearly appreciated if two snapshots corresponding to $0 \: \mu s$ and $600 \: \mu s$ are compared. Also presented in the Fig. \ref{schlieren2} are the mean and standard deviation of pixel intensity values considering $8000$ instantaneous snapshots. The thick and blurred nature of the separation shock in the mean picture also illustrates the unsteady nature of the flowfield. The non-zero values present in the standard deviation picture shown in Fig. \ref{schlieren2}c also demonstrate the same.

\begin{figure*}[htpb]
    \centering
    \includegraphics[width=140mm]{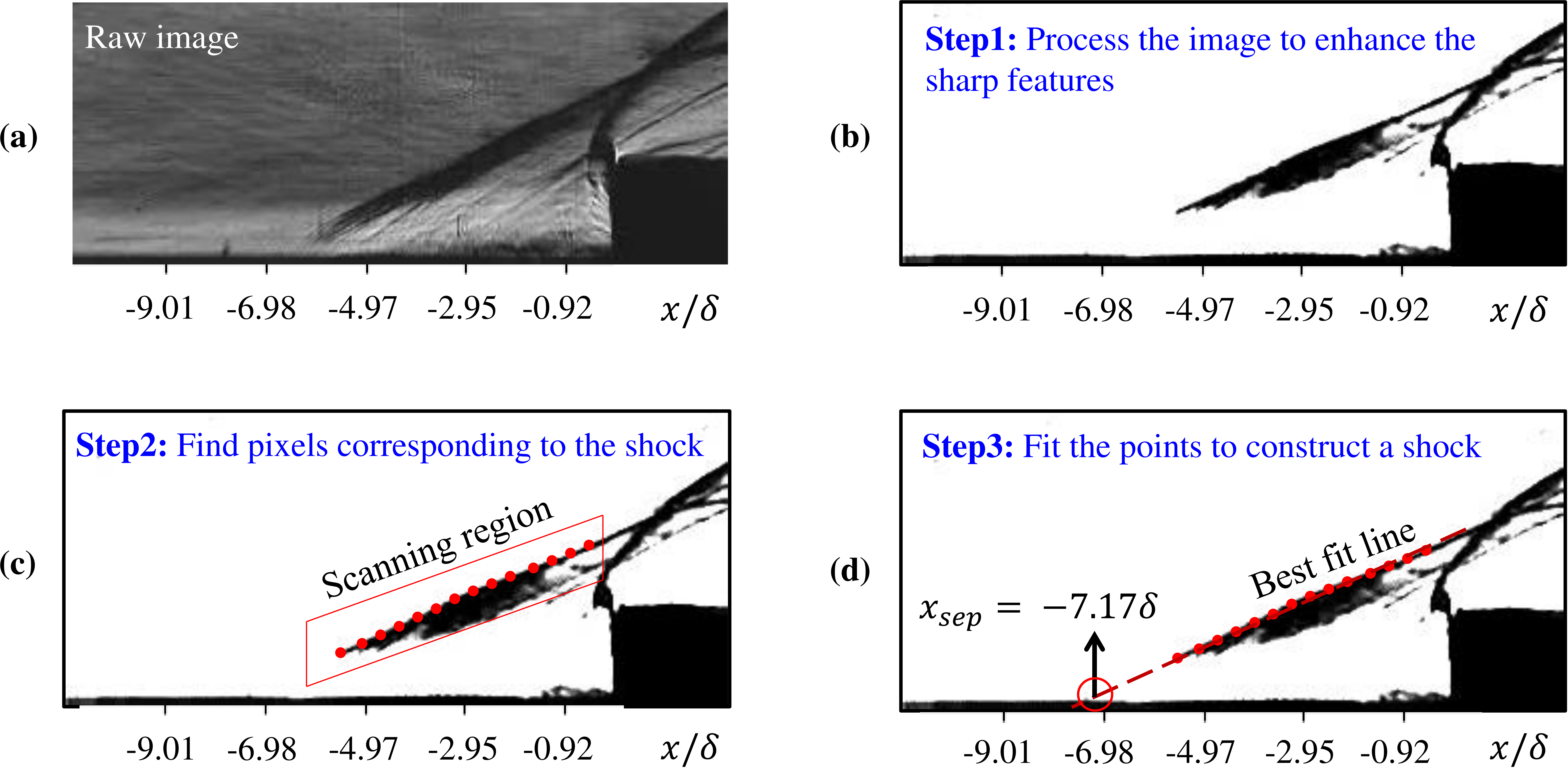}
    \caption{(a) Raw sample snapshot from time resolved schlieren (b) Processed schlieren image with separation shock looking more distinct (c) Scanning region for extraction of pixels corresponding to the separation shock (points on the separation shock indicated). (d) Reconstructed separation shock, using best fit line passing through the first dark pixels from top-down, in the scanning region.}
    \label{shock_extract}
\end{figure*}

To analyze the dynamics of separation shock, the instantaneous separation shock position in the midspan plane, as apparent in the instantaneous schlieren images, was extracted from each snapshot through image processing and a program written to find the pixel locations corresponding to the separation shock. First of all, the image processing was done by adjusting the brightness, contrast and sharpness of the raw image (shown in Fig. \ref{shock_extract}a) so as to display the separation shock with high contrast. This is followed by detecting pixels on the surface of separation shock (in midspan plane), by spotting the locations with a sudden jump in pixel intensity. Subsequently, the shock position was reconstructed as a single straight line in each snapshot through the identified pixel locations. The steps involved in the process are also pictorially shown in Fig. \ref{shock_extract}.

\begin{figure}[htpb]
    \centering
    \includegraphics[width=70mm]{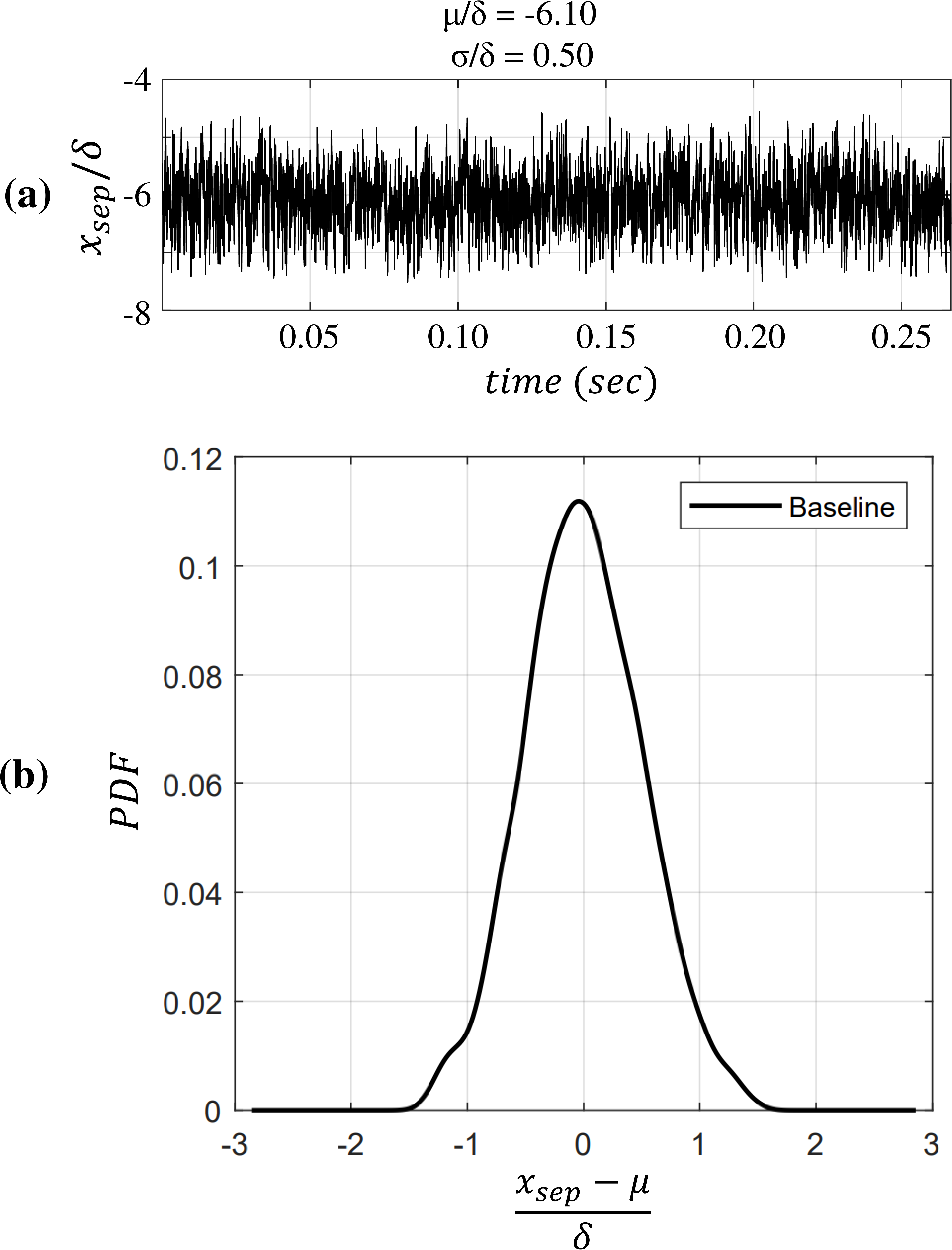}
    \caption{(a) Time series of shock foot collected using image processing. (b) Probability Density Function of shock foot on the spanwise central plane.}
    \label{PDF1}
\end{figure}

The best fit line for the shock is projected onto the wall to obtain the instantaneous shock foot position. After being able to extract shock foot position from each snapshot captured over a time period of $0.27$ sec ($8000$ snapshots), a shock-foot location time series has been plotted (Fig \ref{PDF1}a). The shock foot can be seen to sweep between distances $5\delta$ and $7\delta$ from the front of protrusion (in the spanwise centred plane). The mean position of the shock foot was found to be at $6.1\delta$ from the protrusion. The standard deviation of the shock foot (which represents its degree of presence away from the mean position) was found to be $0.5\delta$. The effect of control on these two quantities will be discussed in the next section with a few important control configurations.

In order to gain an understanding of the degree of shock’s presence at various spatial locations on the bottom surface, the Probability Density Function (PDF) of the shock foot data is obtained. Since the shock-foot data obtained is discrete, kernel density estimation method is used to obtain the PDF (with Gaussian kernel function). The PDF has been plotted about the mean shock foot position. The PDF can be seen to be mostly symmetrical about the mean, with maximum amount probability density in the range of $-\delta$ to $\delta$. This corresponds to a sweeping length of approximately $2\delta$.

%%%%%%% CONTROL section

\section{Effect of suction on the separation and unsteadiness} \label{sec:control1}

A total of ten suction hole configurations were designed and experimented with, by placing the holes only along the midspan in all the layouts. The placement of suction holes were strategically chosen to explore the effect of suction from the three pressure regions identified from surface pressure measurements. The names of the configurations are given in the format C\_no.1\_no.2\_no.3 (first letter C stands for `Configuration'). The numbers: no.1, no.2, etc- denote bleed-hole positions, in terms of their distance (in mm) from the protrusion front-face. For example, C\_2.5\_12.5 represents the configuration with two holes, one of them with its centre at $2.5$ mm from the protrusion, the other with its centre at $12.5$ mm from the protrusion.

% It was hypothesized that the bleed mass flow rate through the vents may play a role in the extent of control achieved. This is envisioned because the extent to which the bleed rate influences near wall flow field is expected to be directly related to the control effect that it induces. 

It was hypothesized that the bleed from different locations on the plate (zones) may result in different bleed flow rates, which in turn can be among the factors determining the control effectiveness of particular configurations. Configuration C\_2.5 employs a bleed hole at 2.5 mm (or $0.36\delta$) from the protrusion, which is in the region of high pressure from where the highest bleed mass-flow is expected. The Other configurations C\_2.5\_12.5, C\_2.5\_12.5\_27.5, and C\_2.5\_27.5 were designed to explore the bleed effect from the hole placed at $x=-2.5$ mm in combination with holes placed in low and plateau pressure regions which correspond to locations $x=-12.5$ mm and $x=-27.5$ mm respectively. The rest of the configurations were designed to explore the control effect due to holes placed solely in the plateau and low pressure regions.

\subsection{Oil flow visualization}

\begin{figure*}[htpb]
    \centering
    \includegraphics[width=140mm]{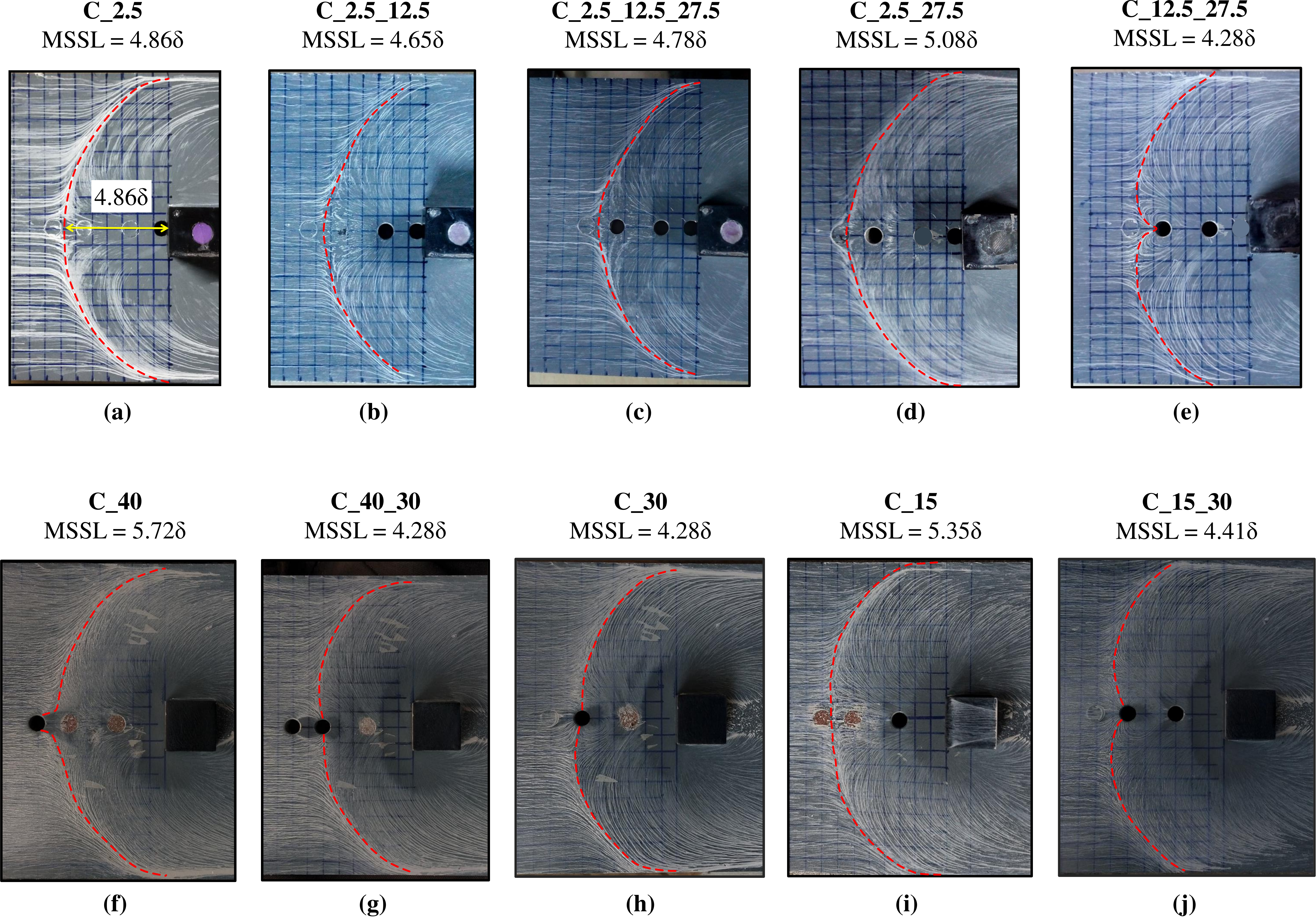}
    \caption{Oil-flow surface streak line patterns for various control cases. The separation line corresponding to the separation shock is indicated with the red dashed line. The configuration name and the Mid Span Separation Length (MSSL) is indicated above each of the corresponding images. Grid lines on the surface of the baseplate can be also seen in the images which are drawn for measurement purposes. Each grid box measures 5 mm ($0.71\delta$) in both its width and height.}
    \label{control-oilflow1}
\end{figure*}

\begin{figure*}[htpb]
    \centering
    \includegraphics[width=145mm]{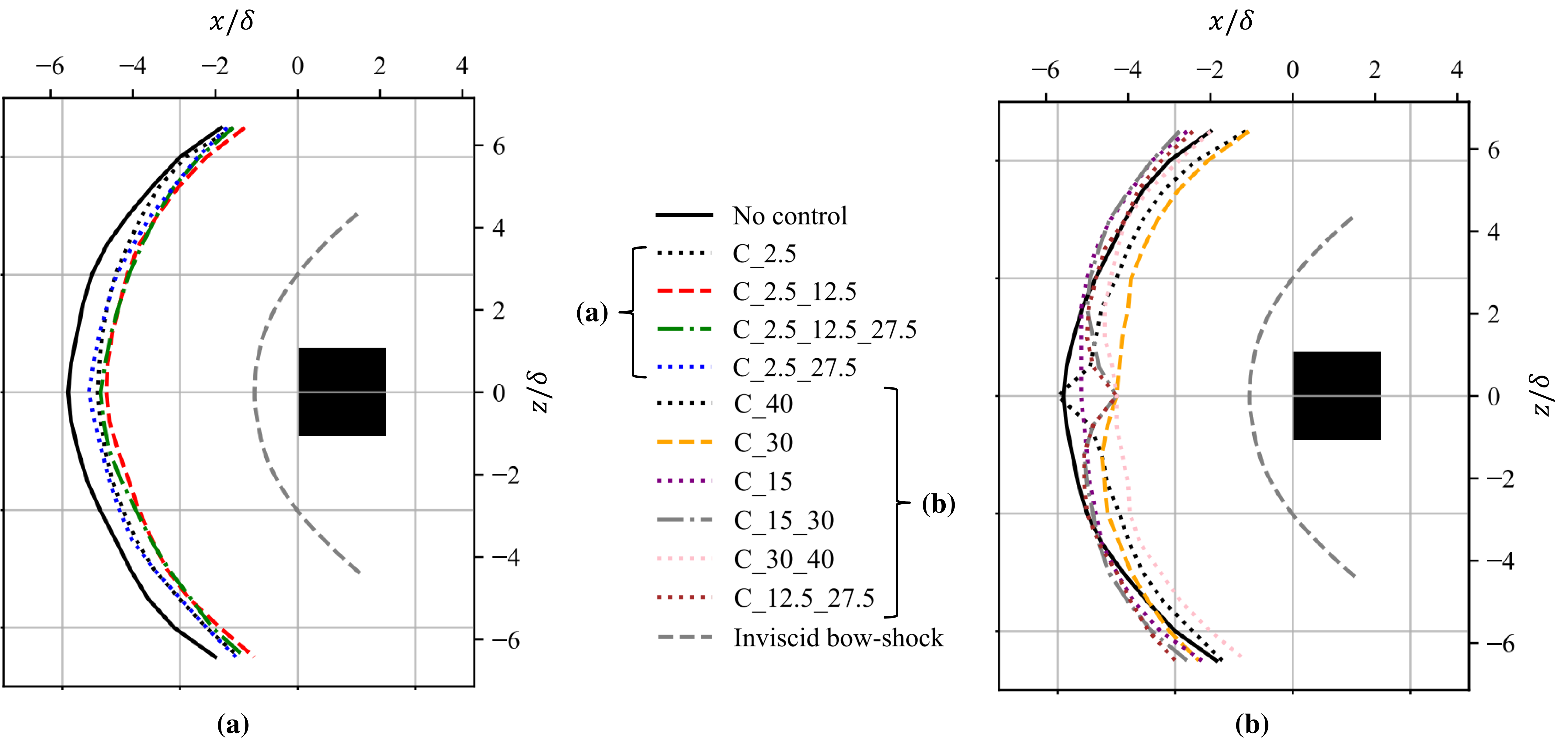}
    \caption{Geometry of separation lines for control configurations (a) C\_2.5, C\_2.5\_12.5, C\_2.5\_12.5\_27.5 and C\_2.5\_27.5 (b) C\_12.5\_27.5, C\_40, C\_40\_30, C\_30, C\_15 and C\_15\_30 in comparison to the geometry of baseline separation line (solid black line).}
    \label{control-oilflow2}
\end{figure*}

The oil-flow streakline patterns generated for various control cases were shown in Fig. \ref{control-oilflow1} along with their corresponding midspan separation lengths. The first four configurations (a,b,c and d) are those with a hole in the high pressure region (at $x=-2.5$ mm), and the rest has suction only from upstream locations. Separation line geometries from these images were extracted with the help of grid markings made on the bottom surface, and plotted together, as shown in the Fig. \ref{control-oilflow2}, to compare the profiles for various cases against the baseline case; inviscid bow-shock profile for the baseline case is also plotted for reference.

At a first glance of the geometry of streakline patterns shown in Fig. \ref{control-oilflow1}, some important features and differences can be noted in comparison with the baseline case. For configurations C\_2.5\_12.5, C\_2.5\_12.5\_27.5 and C\_2.5 a reduction in separation length along the entire span can be clearly observed. Further, for the configurations having a hole in high pressure region ($x=-2.5$ mm), the separation line and the surface streakline pattern were symmetric about midspan. However, for some other configurations the separation line can be seen to be asymmetric (C\_40\_30, C\_30 and C\_15). In some configurations hinging/pulling effect towards the bleed holes can be observed; this hinging of the separation line in the mid-span is followed by the quick catching up of the separation line with that of the baseline case, or even an increase in separation length at some spanwise locations (local separation length being the x-distance from protrusion front face to the separation length at the considered spanwise location). The separation line corresponding to the configuration C\_12.5\_27.5 can be seen to have its position hinged towards the bleed hole at $x=-27.5$ mm ($-3.9\delta$). A similar pulling effect was seen for most configurations (C\_12.5\_27.5, C\_40, C\_40\_30, C\_30) with a bleed hole in the plateau pressure region. It was also noted that the separation line for C\_15\_30 experiencing the pulling effect of suction from the bleed hole at $x = -30$ mm (or $-4.3\delta$), has indeed increased separation length at some locations away from the midspan, although the separation length in the midspan was reduced (with the separation line hinged on the front of the hole). 

To quantify the performance of the control configurations, the Mean Reduction in Separation Length (MRSL) throughout the span was calculated for each configuration. MRSL essentially quantifies the overall amount of reduction/increase in separation length due to the flow control. The values of MRSL for the different control configurations are reported in the Table \ref{MRSL-table}, along with the midspan separation length and bleed mass-flow rate (bleed mass flow was measured only for few critical configurations). The observations presented in the table, in general, suggest that the introduction of suction at the high pressure region will lead to a delay in overall separation. With MRSL of $0.86\delta$, the best performance can be seen for the configuration C\_2.5\_12.5 (combination of two suction holes one from high pressure and another from low). This configuration also had a reduction in midspan separation length of $0.92\delta$ which, however, is not the highest reduction in midspan separation length. Configurations with a hole in the plateau but without a hole in the high-pressure region generally had the highest reduction in midspan separation length, which was $1.29\delta$. However, these configurations had considerably lesser MRSL, one of them (C\_15\_30, with a reduction in midspan separation length of $1.15\delta$) even having a negative MRSL (due to larger separation lengths than baseline case for a wider range of span).

In fact, all the configurations in which the bleed hole was placed at $x = -2.5$mm ($-0.36\delta$) (which is in high pressure zone), can be seen to show good performance in reducing the separation length throughout the span, and also preserving the symmetry about midspan. Further, these configurations have also reported higher bleed mass-flow rates compared to any other configuration without bleed hole at $x = -2.5$ mm. This clearly demonstrates that it is advantageous, in terms of reduction in separation length, to have suction from high pressure region of the separation bubble rather than employing bleed from other places in the midspan.

In configurations C\_15 and C\_15\_30 suction was employed from relatively low pressure regions of the separation region (between $x = -\delta$ to $-2.5\delta$). The performance from these configurations was observed to be poor with an overall increase in the size of the separation bubble. Thus, suction from low pressure regions of the separation bubble can be said to be undesirable in general.

Although the bleed mass-flow that was achieved through configuration C\_2.5 is higher than the configuration C\_2.5\_12.5, control performance was seen to be better for the configuration C\_2.5\_12.5. It can be inferred from this observation that, the bleed placement is of greater importance for flow control than the bleed mass flow rate. In fact, the configuration of bleed placement dictates the bleed mass flow rate.

% \begin{table*}[htpb]
%     \centering
%     \includegraphics[width=100mm]{Images/MRSL-table.pdf}
%     \caption{Comparison of performance and bleed mass-flow rates gained from various control configurations.}
%     \label{MRSL-table}
% \end{table*}

\begin{table*}
\caption{\label{MRSL-table}Comparison of performance and bleed mass-flow rates gained from various control configurations.}
\begin{ruledtabular}
\begin{tabular}{cccc}
Configuration&Midspan separation distance&Mean reduction in separation length (MRSL)&Bleed massflow rate (kg/hr)\\
\hline
C\_2.5\_12.5 & $4.65\delta$ & $0.86\delta$ & 13.1\\
C\_2.5\_12.5\_27.5 & $4.78\delta$ & $0.77\delta$ & -\\
C\_30\_40 & $4.28\delta$ & $0.71\delta$ & 4.35 \\
C\_30 & $4.28\delta$ & $0.68\delta$ & - \\
C\_2.5 & $4.86\delta$ & $0.54\delta$ & $16.6$ \\
C\_2.5\_27.5 & $5.08\delta$ & $0.54\delta$ & $11.9$ \\
C\_40 & $5.72\delta$ & $0.51\delta$ & 3.52 \\
C\_12.5\_27.5 & $4.28\delta$ & $0.02\delta$ & - \\
\textbf{No control} & \boldsymbol{$5.57\delta$} & \boldsymbol{$0$} & \boldsymbol{$0$} \\
C\_15\_30 & $4.41\delta$ & $-0.03\delta$ & - \\
C\_15 & $5.35\delta$ & $-0.06\delta$ & - \\
\end{tabular}
\end{ruledtabular}
\end{table*}

\subsection{Time resolved schlieren visualization}

Time resolved Schlieren visualization data with the control configurations reveals some important details concerning the effect of control on unsteadiness. Experiments were performed on three control cases: C\_2.5\_12.5, C\_2.5 and C\_12.5\_27.5. Configuration C\_2.5\_12.5 was chosen as it is the best performing case of all, and configuration C\_12.5\_27.5 was chosen in order to see the effect of one bleed hole placed close to the separation location but without a hole in high pressure region. Configuration C\_2.5 was chosen to see the solo effect of bleed from the hole in the high pressure region.

\begin{figure}[htpb]
    \centering
    \includegraphics[width=70mm]{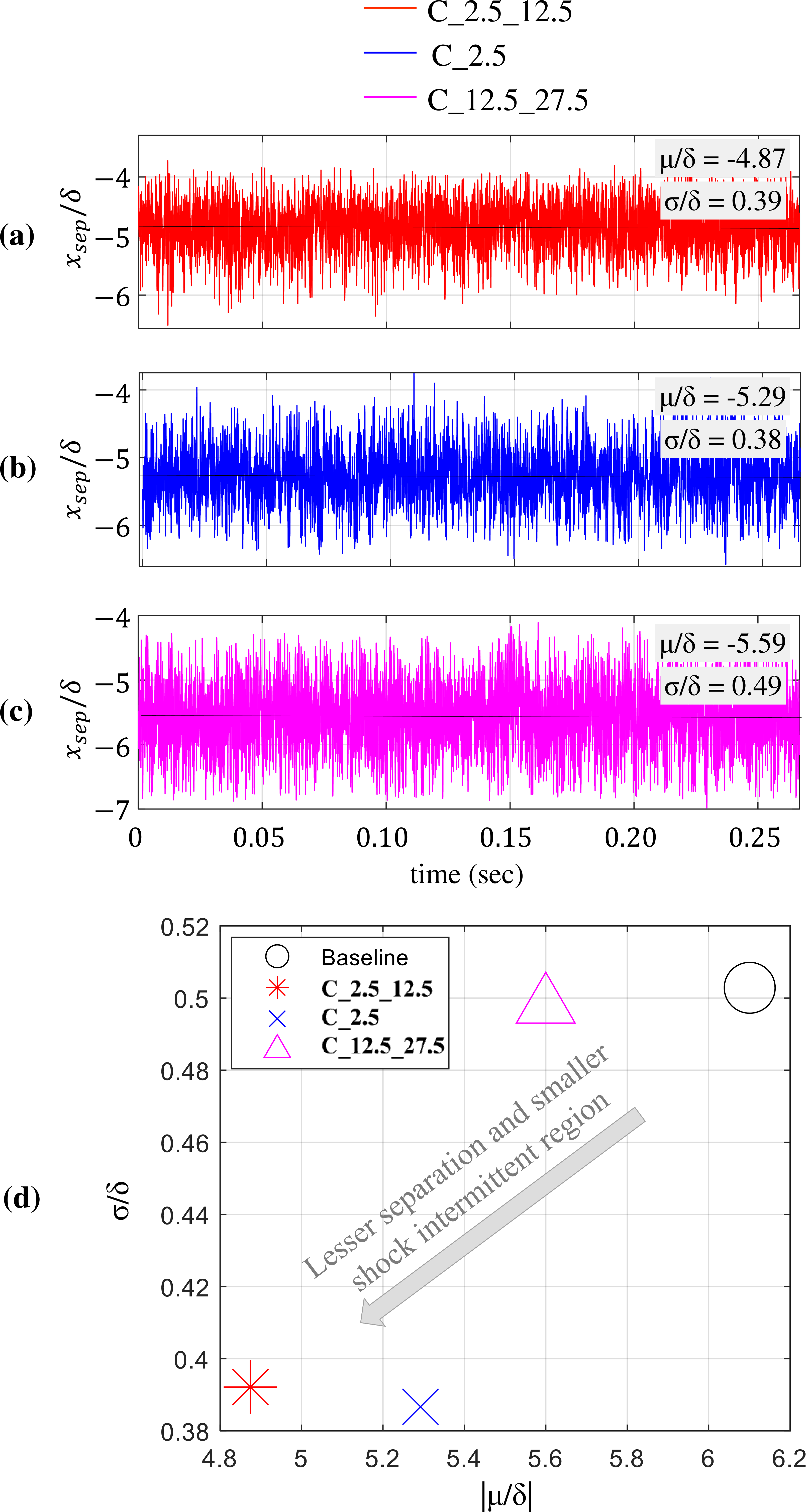}
    \caption{Time series of shock foot for (a) C\_2.5\_12.5, (b) C\_2.5 and (c) C\_12.5\_27.5. (d) Mean separation distance versus standard deviation for baseline and control configurations.}
    \label{control-timeseries}
\end{figure}

The separation shock foot location at the midspan was extracted from each snapshot using the same procedure described in the previous section in Fig. \ref{shock_extract} for the baseline case. The shock foot location time series for control cases C\_2.5\_12.5, C\_2.5 and C\_12.5\_27.5 are shown in Fig. \ref{control-timeseries}(a,b, and c). The corresponding mean and standard deviation of shock foot location for each case was computed.

The mean separation shock location for the control case C\_2.5\_12.5 calculated by averaging the shock foot data is found to be at $4.87\delta$ from the protrusion, which is approximately $20\%$ ($1.23\delta$) lesser (downstream) than the baseline case’s mean separation length ($6.1\delta$). In a similar fashion, the mean shock foot distances for the cases C\_2.5 and C\_12.5\_27.5 are also lower by $13\%$ and $8\%$ each with their mean positions at distances $5.29\delta$ and $5.59\delta$ from protrusion foot respectively. One can note that for the C\_12.5\_27.5 case, although the separation point at midspan is much downstream ($4.28\delta$), the reduction in midspan separation shock is not as much as the reduction in midspan separation length. It must also be noted that this configuration almost has zero MRSL, and at midspan the separation line turns sharply and reaches the position of the baseline case in a short spanwise distance, which could be the reason for the observed mean shock foot position at midspan.

A plot of Mean ($\mu / \delta$) versus Standard deviation ($\sigma/\delta$) of shock foot position obtained for each of these cases is shown in Fig. \ref{control-timeseries}d to understand the relation between the effect of control in reducing `mean separation length' and in `shock intermittent region (sweeping region)'. The arrow in the image indicates the direction towards a desirable flowfield condition. Broadly, suppression in the intermittent region seems to be associated with a reduction in overall separation length. It can be noted that despite having only one bleed hole in the case of C\_2.5 it has performed significantly well, especially in suppressing the shock intermittent region. In addition to the bleed hole in the high pressure region, when a bleed hole is also introduced in the low pressure region (C\_2.5\_12.5), the combined performance of suction from these two holes results in nearly comparable suppression of shock intermittent region. This is also the configuration with the best MRSL. This clearly demonstrates the superior effectiveness of bleed from high-pressure region in controlling both separation and shock oscillations. However, with suction from one hole in low pressure region and one in the plateau region (i.e., for C\_12.5\_27.5 case) the performance does not look the best in terms of reduction in both separation length and standard deviation. 

\begin{figure}[htpb]
    \centering
    \includegraphics[width=70mm]{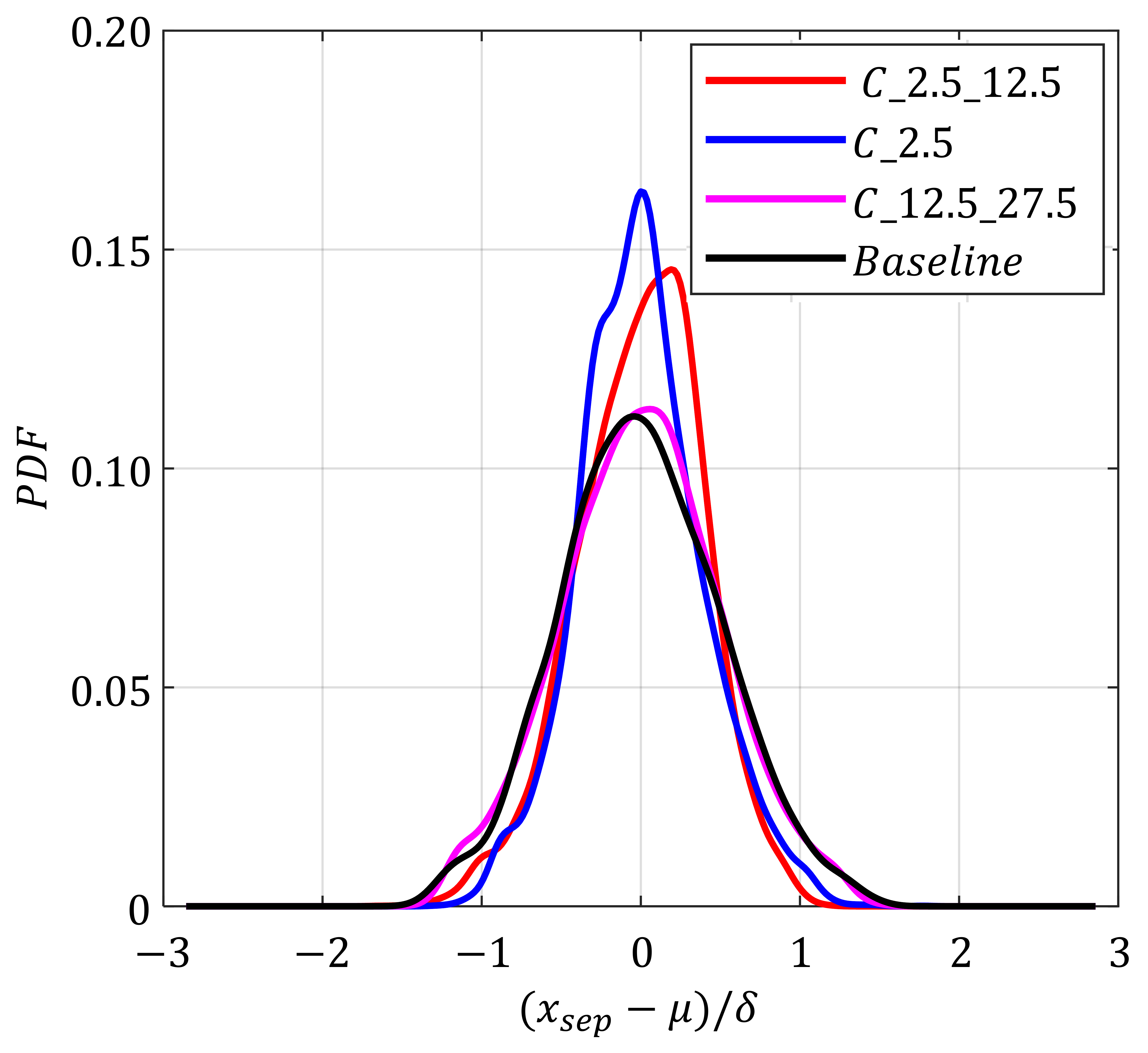}
    \caption{Probability Density Functions (PDFs) of shock foot time series about mean for baseline, C\_2.5, C\_2.5\_12.5 and C\_12.5\_27.5 control configurations}
    \label{control-PDF}
\end{figure}

Probability Density Function (PDF) plots were obtained for the control cases too based on the shock foot data presented in Fig \ref{control-timeseries}(a,b, and c), using the same procedure employed for the baseline case as discussed in section \ref{sec:schlieren1}. Fig \ref{control-PDF} shows probability density of control cases C\_2.5\_12.5, C\_2.5 and C\_12.5\_27.5 along with the baseline case. All the curves can be seen to be symmetric with Gaussian like distribution. The curves corresponding to the configuration C\_2.5\_12.5 and C\_2.5 are narrower and taller relatively, which illustrates that the standard deviation corresponding to these two cases is considerably low when compared with the other two cases. This also implies that the intermittent region corresponding to these two configurations is comparatively smaller. However, the distributions for C\_12.5\_27.5 and baseline cases are more spread out and have relatively lower peak values. This corresponds to the fact that their separation shock foot oscillates in a wider region.

%%%% LAST SECTION before conclusions 

\section{Fourier and Proper Orthogonal Decomposition (POD) analysis of shock unsteadiness} \label{sec:control2}

The reduction in mean separation lengths and shock foot sweeping range, with the introduction of suction control in configurations C\_2.5\_12.5 and C\_2.5 was clearly recognized from the statistical analysis presented in the previous subsection. In this subsection, the frequency components present in the shock oscillations for baseline, C\_2.5\_12.5 and C\_12.5\_27.5 cases are analysed by performing Fourier analysis and Proper Orthogonal Decomposition (POD) of temporal schlieren snapshot data. 

\subsection{Fourier analysis of shock oscillations} \label{Fourier}

Time series of shock foot position was obtained from the best line fit for the separation shock from the instantaneous snapshots as discussed in the previous section. However, since the best line fit could filter out some frequencies which could otherwise be observed at individual shock locations, shock foot time series is not subject to Fourier analysis. Instead, a scan of intensity along a horizontal scanning line passing through the shock, at the level of the edge of the (undisturbed) boundary layer, is used to identify the instantaneous shock position at this particular height, whose time series was analysed.

\begin{figure}[htpb]
    \centering
    \includegraphics[width=70mm]{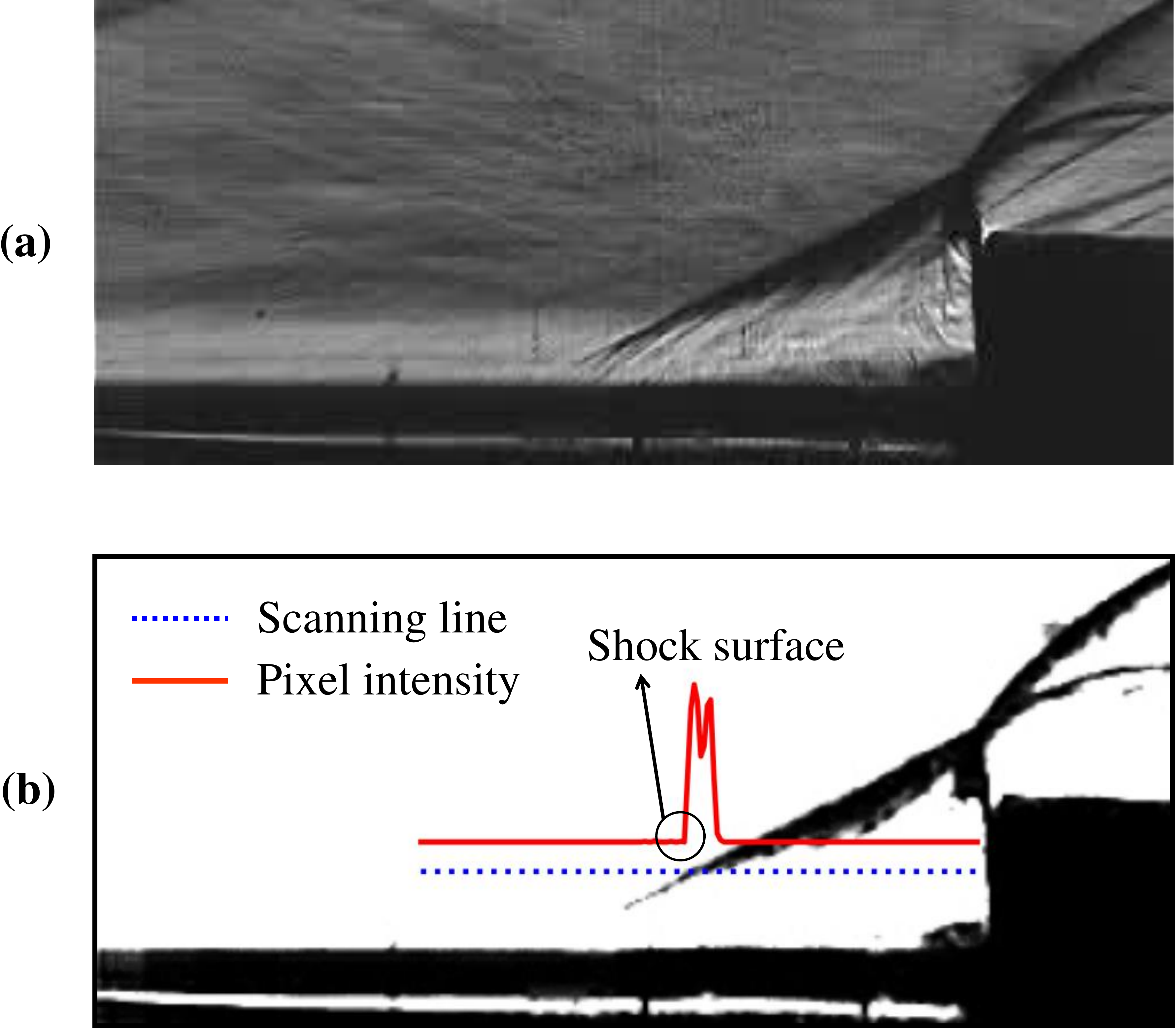}
    \caption{(a) Sample instantaneous raw schlieren image, (b) Processed schlieren image with scanning line and intensity along the line indicated for shock position detection.}
    \label{pixel_scan}
\end{figure}

The schlieren snapshots were first processed before extracting the shock position information to increase the precision of the analysis. Fig. \ref{pixel_scan} shows a processed instantaneous schlieren image acquired using the procedure described in section \ref{sec:schlieren1} (with reference to Fig. \ref{shock_extract}). The shock was detected by scanning pixel intensity values from left to right along a scanning line at $y=\delta$ shown with blue dotted lines in Fig \ref{pixel_scan}. The corresponding intensity variation along the scan line has also been shown above it with a solid red line. The pixel location corresponding to the first sudden change in intensity value was taken as the position of the separation shock surface in the midspan passing through the scan line in each snapshot. The process was automated through a python program to locate shock position on the scan line in all other snapshots for baseline, C\_2.5\_12.5, and C\_12.5\_127.5 cases. The time series containing locations of pixels corresponding to separation shock on the scan line for each snapshot for various cases is obtained and used for the Fourier analysis, whose results are presented subsequently.

% The \texttt{pwelch} function in MATLAB was used to estimate the Power Spectral Density (PSD) - $G(f)$ in the Fourier analysis presented in this paper.

\begin{figure}[htpb]
    \centering
    \includegraphics[width=70mm]{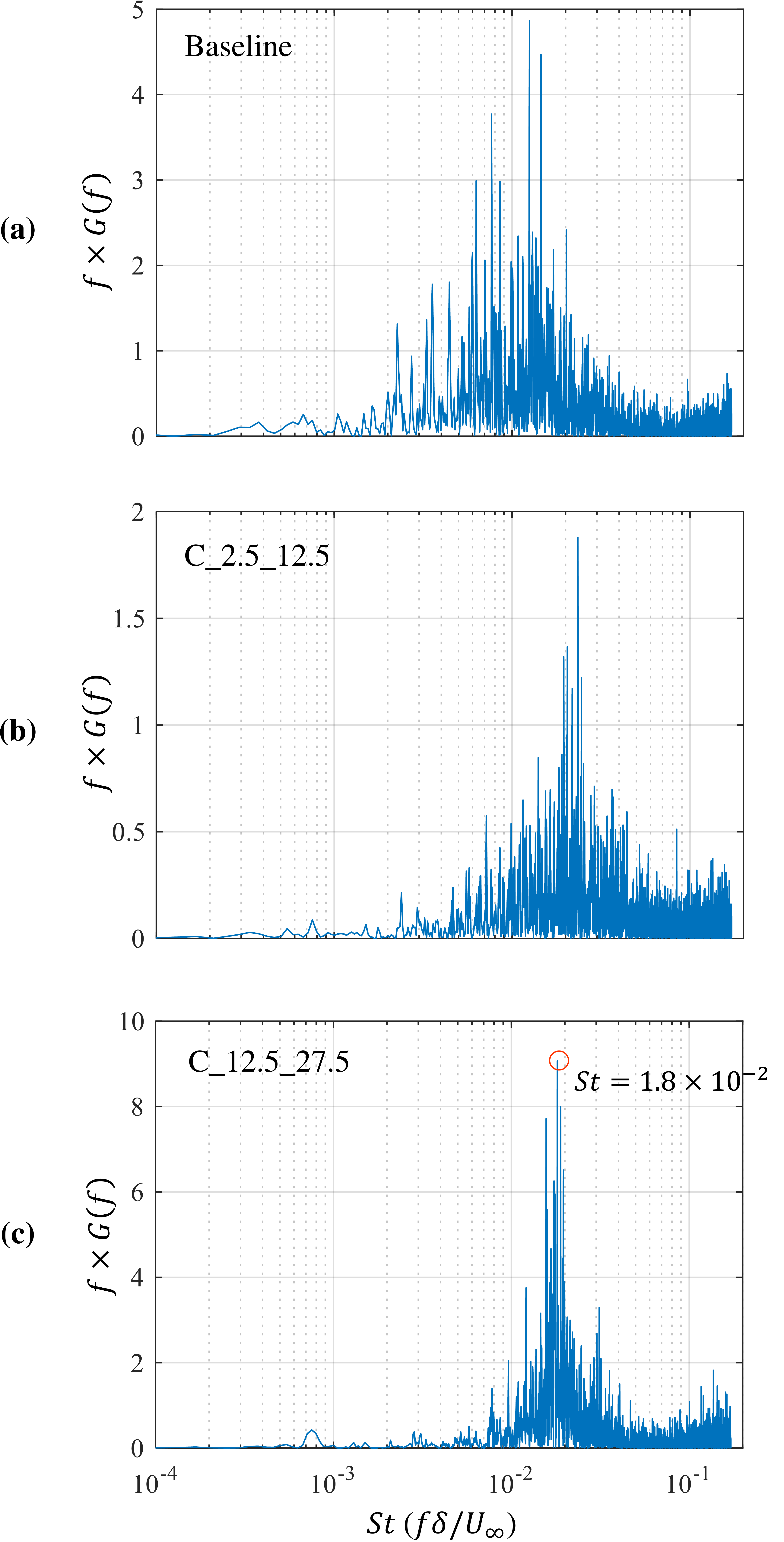}
    \caption{Power spectral density (PSD) of shock position extracted from schlieren images for: (a) Baseline, (b) C\_2.5\_12.5 and (c) C\_12.5\_27.5 cases.}
    \label{PSD1}
\end{figure}

To estimate the Power Spectral Density (PSD) - $G(f)$ for the Fourier analysis presented in this section, MATLAB's \texttt{pwelch} function was used. A total of $8000$ snapshots captured at $30,000$ Hz was used for the analysis, which suffices to resolve the expected low-frequency unsteadiness (around $1000$ Hz) characteristic of two-dimensional SBLIs. Fig. \ref{PSD1} shows the pre-multiplied Power Spectral Density ($f \times G(f)$) versus the Strouhal number for time-series data constructed for baseline, C\_2.5\_12.5 and C\_12.5\_27.5 cases using the procedure mentioned above. While it is observed that the most energetic Strouhal numbers for all three cases are in the order of $10^{-2}$, a difference in the spread of the dominant frequencies is noted among them. These frequencies, which correspond to three-dimensional interactions, can be noted to be the same as the widely reported low frequencies found in two-dimensional interactions\cite{clemens2014low}. In comparison to the baseline case, the spectrum of dominant oscillation frequencies corresponding to C\_2.5\_12.5 is spread widely around $St = 10^{-2}$ with lesser power. This suggests that the effect of suction has resulted in wider distribution of energy among various frequencies. In contrast to this, configuration C\_12.5\_27.5 has exhibited a distinct peak at a Strouhal number of $1.8\times10^{-2}$ ($\approx 1600$ Hz). This interesting observation suggests that the presence of a bleed hole at the separation zone causes separation shock to oscillate almost periodically. Further, it may be recalled that asymmetry about mid-span was observed in separation lines for some cases with upstream bleed holes (such as separation/plateau pressure zones). It leads to a speculation that these `oscillations' could also be characterised by alternating forward and backward motion of the separation line on either side of the mid-span, although a confirmation of this speculation requires sophisticated diagnostics and visualisation.

% This might be due to the presence of a bleed hole at the separation zone, which causes the separation shock to oscillate at a specific frequency.

\subsection{Snapshot POD analysis}

Proper Orthogonal Decomposition (POD) is a dimensionality reduction technique that can be used to decompose temporal flow-field data into a series of respective “orthogonal spatial basis functions” or also called as “POD modes”. Each POD mode presents a low-dimensional description of the entire unsteady flow-field with certain energy. This technique has been widely used by fluid dynamicists since 1967 \cite{lumley1967structure} especially on the experimental PIV data-sets \cite{fujiwara2020experimental} and data from computational simulations \cite{sengupta2004proper} to uncover important spatial and temporal scales associated with the flow-field. However, in a few recent studies, the technique has also been applied to the field of view schlieren snapshots \cite{berry2017application}. In addition, the temporal characteristics associated with each mode can also be studied by performing Fourier analysis of ‘temporal mode coefficients’ which are obtained by projecting the snapshot vector data onto the intended orthogonal basis vector (POD mode). These ‘temporal mode coefficients’ essentially carry dynamical information corresponding to the mode. Details of the various steps and the algorithm used to perform snapshot POD in the current work are being omitted here for brevity. A good summary of snapshot POD and the method used in the current study can be found in \citet{meyer2008identify}.

The current subsection is dedicated to extract frequencies corresponding to separation shock dynamics as observed in dominant POD modes and compare them with the frequencies obtained in the previous sub-section \ref{Fourier} where the separation shock is tracked using image processing.  To do so, various mode shapes were extracted from the time resolved schlieren data, and Fourier analysis was done on the mode coefficients to extract the frequencies present in each corresponding mode. Before proceeding into the POD analysis of shock oscillations, firstly, the outlier frequencies that do not contribute to the flow physics will be identified. 

\subsubsection{Identification of the outlier frequencies}

\begin{figure}[htpb]
    \centering
    \includegraphics[width=80mm]{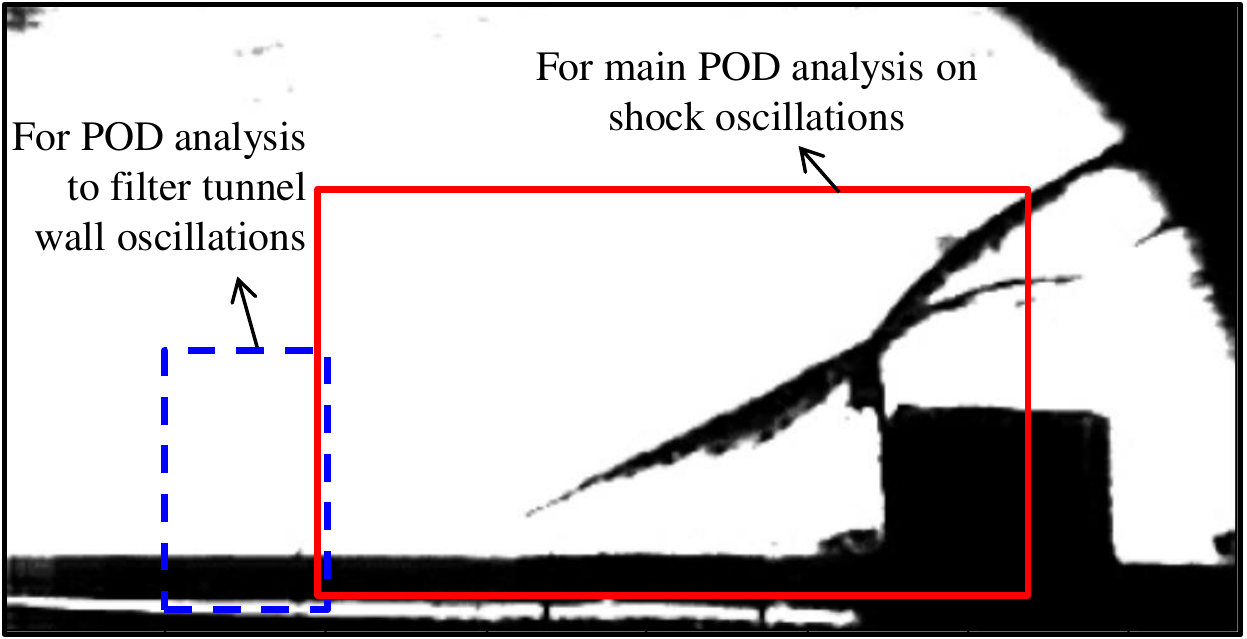}
    \caption{A processed schlieren snapshot and windows of cropped regions considered for POD analysis.}
    \label{POD_windows}
\end{figure}

\begin{figure}[htpb]
    \centering
    \includegraphics[width=85mm]{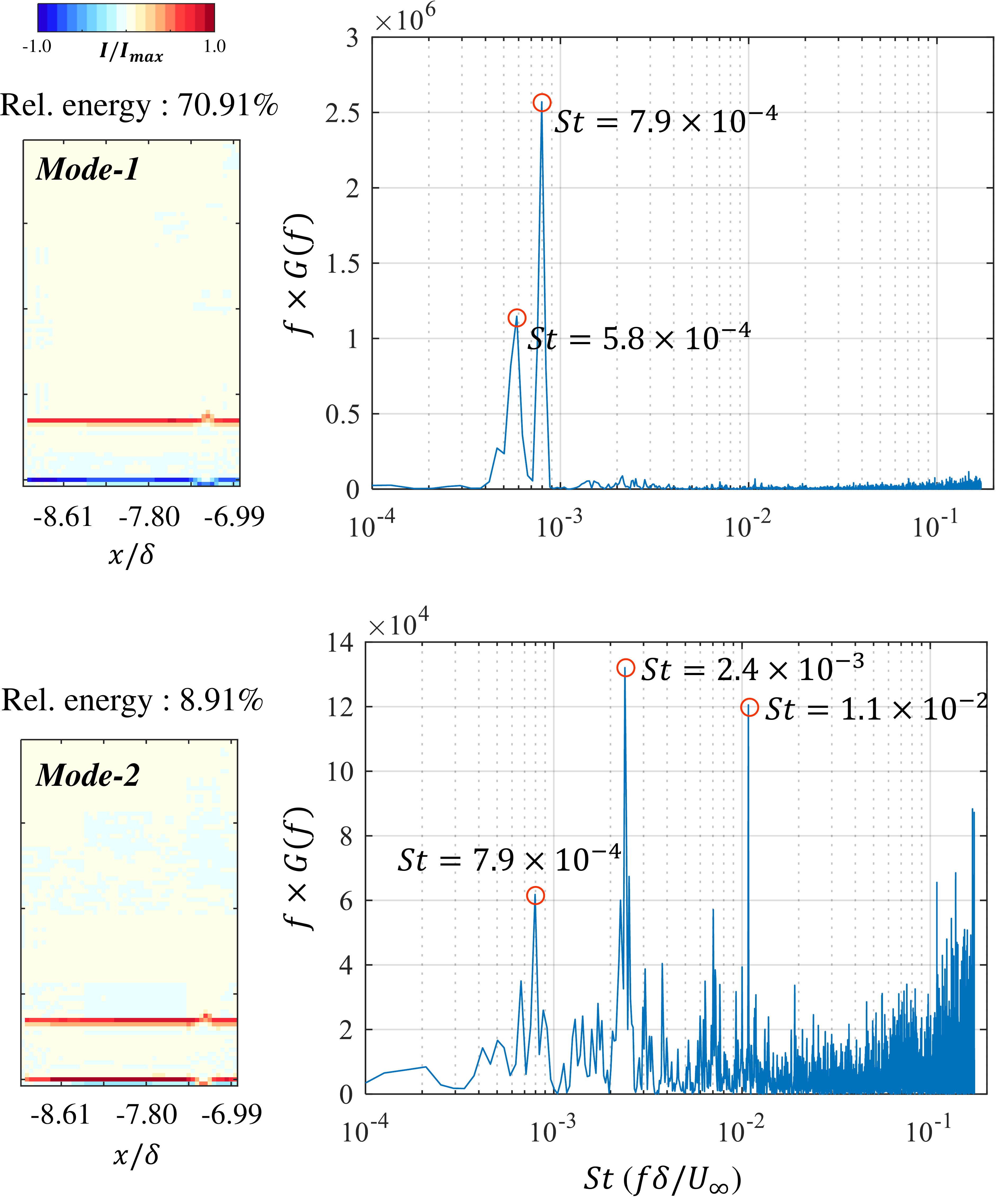}
    \caption{(a) Various POD mode structures (corresponding mode and energy is indicated on top) and (b) PSD spectrum of POD mode coefficients.}
    \label{POD_wall}
\end{figure}

To avoid any frequencies corresponding to freestream turbulence to appear in the spectrum, POD analysis was done on the processed images which were obtained through the method detailed in section \ref{sec:schlieren1} (with reference to Fig. \ref{shock_extract}). Further, the analysis of the SBLI was only restricted to a cropped portion of the entire image, marked with a red box in Fig. \ref{POD_windows}. Before carrying out the main POD analysis, the outlier frequencies such as the low frequencies corresponding to the tunnel/set-up vibrations were characterized by performing a separate POD analysis. This POD analysis was done on a window containing only the wall and freestream which is highlighted with a blue dashed rectangle in Fig. \ref{POD_windows}. The mode-1 of the analysis, shown in Fig. \ref{POD_wall}a, which carries $70.91\%$ of the total energy, clearly corresponds to tunnel wall motion; this can be inferred by noting the highlighted portions in the mode shape. The low frequencies corresponding to this mode are of Strouhal numbers $5.8\times10^{-4}$ and $7.9\times10^{-4}$ which needs to be dis-regarded if appeared in the POD analysis done in the main window. The second mode shown in Fig. \ref{POD_wall}b carries much lesser energy ($\approx9\%$) in comparison with the first mode, and can be neglected.

\subsubsection{POD modes and the spectrum of temporal coefficients}

\begin{figure*}[htpb]
    \centering
    \includegraphics[width=160mm]{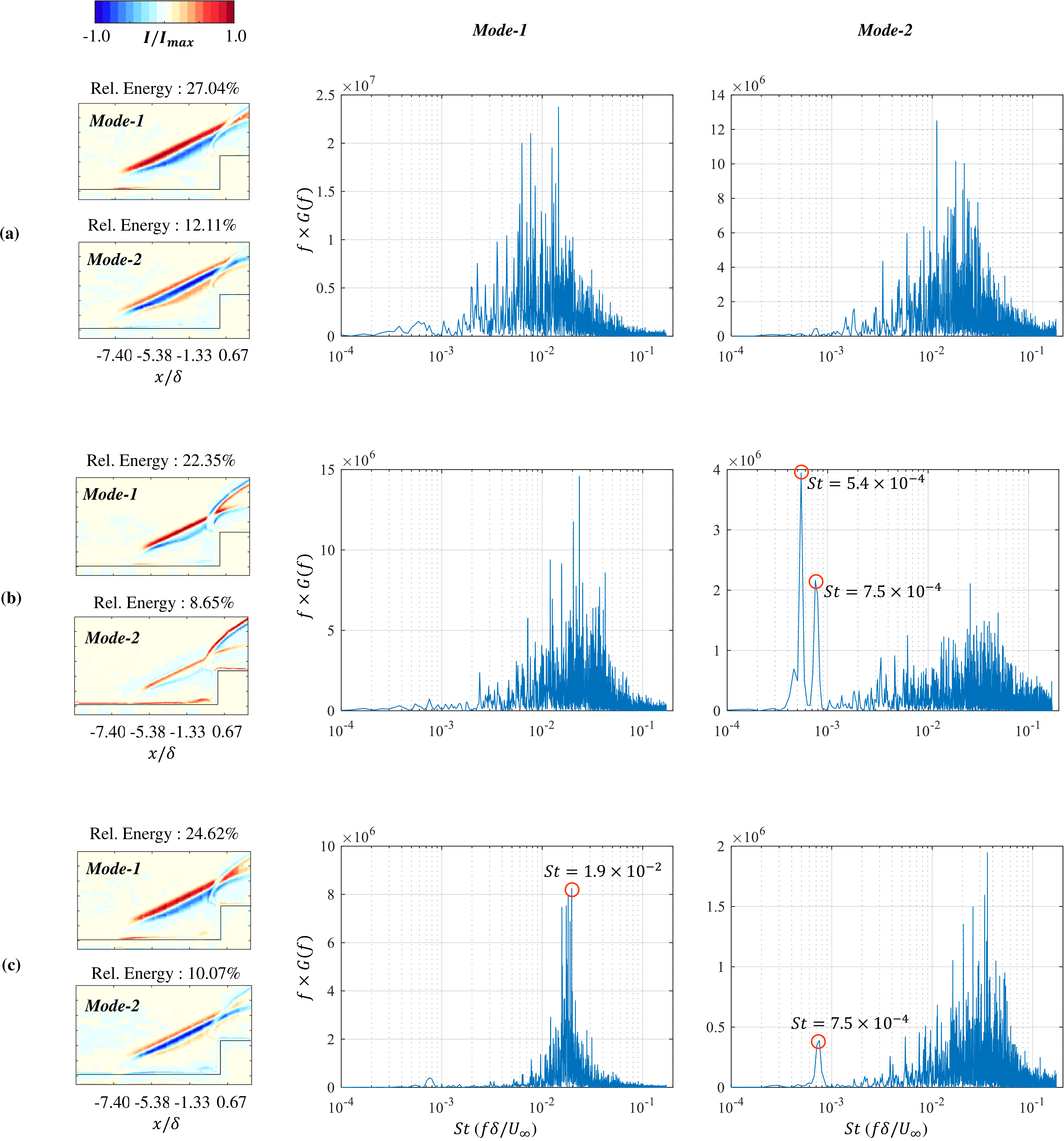}
    \caption{Mode shapes and PSD of mode coefficents for (a) Baseline (b) C\_2.5\_12.5 (c) C\_12.5\_27.5. The mode number and its energy are indicated on top of each mode shape.}
    \label{POD-SBLI}
\end{figure*}

The first two decomposed POD modes of schlieren data (density gradient fluctuation fields) and their corresponding spectrum are presented in Fig. \ref{POD-SBLI} for baseline, C\_2.5\_12.5, and C\_12.5\_27.5 cases respectively. The corresponding mode energy is also presented at the top of each mode shape. The color range for the intensity of pixels ($I$) in the mode images presented here is normalized with the maximum intensity value ($I_{max}$) in the image. It was observed that, in all three cases, a significant amount of energy ($>30\%$) is present in the first two modes. In addition, a drastic drop in energy content can be observed between the first and second modes in all three cases. It can be observed that the region corresponding to the separation shock motion is clearly highlighted in the first mode structure in all three cases. This means that physically the first mode carries the dynamics corresponding to separation shock oscillations in all three cases (Fig \ref{POD-SBLI}a,b, and c). The contribution from other parts of the flow is noted to be minimal in mode-1 as the magnitude of mode intensity values at locations apart from the separation shock region are close to zero. In comparison to mode-1 of the baseline case, the high and low-intensity regions (red and blue) of C\_2.5\_12.5 mode-1 shown in Fig. \ref{POD-SBLI}b can be noted to be much thinner. It can be re-confirmed here again that for case C\_2.5\_12.5, the intermittent region corresponding to the separation shock is thinner and oscillation amplitudes are much less.

The spectrum of mode coefficients for the first two modes of the baseline case and C\_2.5\_12.5 case can be noted to have frequencies that are distributed in a broadband range around $St=10^{-2}$. Whereas in the case of C\_12.5\_27.5, the mode-1 spectrum has a relatively narrow band distribution with high amplitudes peaks close to $St=1.9\times 10^{-2}$ or $f=1660$ Hz (indicated in the figure with a red circle). These distinct peaks in the narrow band suggest a possible periodic oscillation of the shock for this bleed configuration, as also noted in the previous subsection on shock oscillations. The observations noted here in the Strouhal number values for both baseline and control cases are in great consistency with the spectra shown in Fig. \ref{PSD1}.

It can be noted from the PSD plots that while the frequencies have slightly increased due to the introduction of control, the amplitudes have reduced almost by a factor of two. For instance, while the peak amplitudes corresponding to mode-1 of the baseline case is close to $2 \times 10^{7}$, the peak amplitudes for the mode-1 spectrum of control cases are only close to $1 \times 10^{7}$. The lower amplitudes may suggest lesser shock sweeping lengths, which is an indication of the amount of control achieved. In the mode-2 spectrum of C\_2.5\_12.5 and C\_12.5\_27.5 cases, the tunnel oscillation frequencies are clearly visible. These Strouhal number values (marked in the mode-2 spectra of Fig. \ref{POD-SBLI}) are in very close agreement with the frequencies identified in Fig. \ref{POD_wall}a. Considering the amplitudes of tunnel oscillations as a reference, the energy of oscillations corresponding to SBLI in C\_2.5\_12.5 case is much lower and comparable with those of tunnel oscillations, unlike the baseline case, which has much higher oscillation energy with reference to tunnel oscillations.

To summarize the observations presented in this section, firstly, it was noted that the introduction of suction at high and low-pressure regions together caused the shock oscillation amplitudes to reduce by almost two times. Consequently, this has caused the separation shock to oscillate in a smaller intermittent region with approximately two times higher frequency compared to the baseline case. However, when the suction was introduced at the low and plateau pressure regions, the oscillation amplitudes (in mode-1 Fig. \ref{POD-SBLI}c) and the intermittent region's size remained almost unchanged. Nevertheless, the frequency distribution of the shock oscillations has turned to a narrowband, with shock oscillating at approximately two times the peak frequency noted in the baseline case. These observations, in general, suggest that suction from the deeper part of the separation bubble, i.e., from the high-pressure zone offers a better control strategy.

\section{Conclusions}
Experiments with supersonic flow ($M = 2.87$) over square protrusion, roughly twice the local boundary layer thickness ($\delta$) in width and height, resolved some critical aspects associated with the three-dimensional shock structure and separation bubble in-front of the protrusion. Separation line was identified using surface oil flow visualization; the separation at spanwise centre was identified to be $5.57\delta$ from the protrusion face. From the time-resolved schlieren observations, the separation shock-foot was found to be at $6.1\delta$ from protrusion on an average, and had a shock sweeping-length of approximately $2\delta$. It was noted from the surface pressure survey that the three-dimensional relieving effect, observable from the streaklines from oil flow visualization, has resulted in three pressure zones, namely: plateau region downstream of separation, followed by a low-pressure region, and then a high-pressure region in front of the protrusion, unlike only two pressure zones that are typically found in two-dimensional interactions. The measurements and insights obtained from these experiments provided the basis for the design of bleed configurations (in positioning the suction holes at various pressure zones along the spanwise centreline), and served as the baseline case with which the experimental results with bleed control were compared.

Experiments with bleed control were performed with a total of ten different suction hole configurations. Oil-flow visualizations have revealed that the effect of suction from various pressure regions has different implications on the separation line geometry, and the unsteady dynamics. A few configurations such as C\_2.5\_12.5 and C\_2.5\_12.5\_27.5 with a bleed hole in the high pressure region have showed reduction in separation length throughout the span, along with midspan reduction of $0.92\delta$ (16.5\% less) and $0.80\delta$ (14.2\% less) respectively. On the other hand, configurations without any hole in high pressure region and with a hole in plateau region, such as C\_12.5\_27.5 and C\_15\_30 had only shown a reduction in midspan separation length (as high as $1.29\delta$ in some cases), due to the presence of bleed hole close to the separation line. Despite smaller separation lengths along the midspan in these two cases, the overall performance in reducing separation length along span was observed to be poor, due to the larger separation length when compared with baseline case in some range of spanwise distances away from spanwise center. To quantify the overall reduction in separation length across span, Mean Reduction in Separation length (MRSL) was computed as the spanwise averaged distance of the separation line from the protrusion face. Out of all the configurations employed, configuration C\_2.5\_12.5, having two bleed holes, one in the high-pressure zone and other in low-pressure zone, showed good overall performance, with MRSL of $0.86\delta$. However, in few configurations without any hole in the high pressure zone, the MRSL was even observed to be negative, that is, having an increase in separation length overall despite higher reduction in separation length in midspan. Comparison of observations from various control configurations suggests that bleeding the flow from the high pressure region in the separation bubble (near the foot of the protrusion) will produce high reduction in the separation lengths in general. The bleed mass-flow rate measurements suggest that the control performance has a stronger dependency on the hole placement rather than the net amount of bleed rate through the holes.

The Fourier analysis and snapshot POD has illustrated the low frequency separation shock unsteadiness (similar to two-dimensional interactions) in the order of $St = 10^{-2}$ for baseline and control configurations C\_2.5\_12.5 and C\_12.5\_27.5. Specifically, broadband frequencies in the range of approximately $St = 3 \times 10^{-3}$ to $4 \times 10^{-2}$ ($250$ to $3500$ Hz) were observed in baseline and C\_2.5\_12.5 cases. Along with the reduction in mean separation length and shock sweeping length discussed above, the introduction of suction in control configuration C\_2.5\_12.5 has also attenuated the amplitude of frequency peaks observed in the spectrum meaning the oscillations are comparatively less energetic. In contrast to the baseline and C\_2.5\_12.5 cases, a narrow band, almost looking like a single peak, around a $St = 1.9 \times 10^{-2}$ ($\approx 1600$ Hz) was noted for C\_12.5\_27.5 case, with the bleed hole placed near the separation line, making the oscillations periodic.

\begin{acknowledgments}
The research reported in this article is partly supported by the Science and Engineering Research Board (SERB), New Delhi, under Grant No. SRG/2019/001793. The authors would like to thank Prof. G Rajesh and his group at the Department of Aerospace Engineering, IIT Madras, for kindly letting us use some of their experimental equipment for this study. Additionally, the authors would like to thank Siva Vayala for the discussions on this topic.
\end{acknowledgments}

\bibliography{aipsamp}

\end{document}